\documentclass[final,5p,times,twocolumn]{elsarticle}

\usepackage{graphicx}
\usepackage{amssymb,amsmath,bm}
\usepackage{amsthm}
\usepackage{txfonts}
\usepackage{color}
\PassOptionsToPackage{hyphens}{url}\usepackage{hyperref}

\newcommand{\rd}{\mathrm{d}}
\newcommand{\re}{\mathrm{e}}

\journal{Journal of Non-Newtonian Fluid Mechanics}

\begin{document}

\begin{frontmatter}


\title{Non-Newtonian fluid--structure interactions: Static response of a microchannel due to internal flow of a power-law fluid}



\author[purdue]{Vishal Anand}
\ead{anand32@purdue.edu}

\author[iitm,purdue]{Joshua David JR}

\author[purdue]{Ivan C.\ Christov\corref{corrauth}}
\ead{christov@purdue.edu}
\ead[url]{http://tmnt-lab.org}

\cortext[corrauth]{Author to whom correspondence should be addressed.}

\address[purdue]{School of Mechanical Engineering, Purdue University, West Lafayette, Indiana 47907, USA}

\address[iitm]{Department of Ocean Engineering, Indian Institute of Technology Madras, Chennai 600036, India}

\begin{abstract}
We study fluid--structure interactions (FSIs) in a long and shallow microchannel, conveying a non-Newtonian fluid, at steady state. The microchannel has a linearly elastic and compliant top wall, while its three other walls are rigid. The fluid flowing inside the microchannel has a shear-dependent viscosity described by the power-law rheological model. We employ lubrication theory to solve for the flow problem inside the long and shallow microchannel. For the structural problem, we employ two plate theories, namely Kirchhoff--Love theory of thin plates and Reissner--Mindlin first-order shear deformation theory. The hydrodynamic pressure couples the flow and deformation problem by acting as a distributed load onto the soft top wall. Within our perturbative (lubrication theory) approach, we determine the relationship between flow rate and the pressure gradient, which is a nonlinear first-order ordinary differential equation for the pressure. From the solution of this differential equation, all other quantities of interest in non-Newtonian microchannel FSIs follow. Through illustrative examples, we show the effect of FSI coupling strength and the plate thickness on the pressure drop across the microchannel. Through direct numerical simulation of non-Newtonian microchannel FSIs using commercial computational engineering tools, we benchmark the prediction from our mathematical prediction for the flow rate--pressure drop relation and the structural deformation profile of the top wall. In doing so, we also establish the limits of applicability of our perturbative theory.
\end{abstract}



\end{frontmatter}


\section{Introduction}
\label{sec:intro}

A \emph{non-Newtonian} fluid is any fluid for which the shear stress is not simply proportional to the rate of shear strain \cite{CR08,C10}. With the notable exception of air and water, this definition suggests that most fluids, especially in industrial applications, are non-Newtonian \cite{CR08}. Examples include foams, suspensions, slurries, polymeric mixtures, amongst others \cite{BAH87}. In particular, these \emph{complex} fluids are encountered in chemical processing applications in the form of dyes, cements, printing inks, adhesives, artificial food flavors, etc.\ \cite{CR08}. The non-Newtonian  behavior of fluids at the {micron scale (which, in the present context, is still well into the continuum regime)} is attributed, in many cases, to their microstructure. For example, biomolecules such as DNA and proteins suspended in a solution (Newtonian fluid) lead to non-Newtonian response of the mixture due to the chain-like constituents interacting under flow \cite{Anna13}. 

Non-Newtonian fluid mechanics at the {micron scale} is an emerging field due to the potentially new physics exhibited by non-Newtonian fluids in the augmented Weissenberg versus Reynolds number parameter space \cite{GR14}. For example, the limitations imposed by macroscale mechanical devices for fluid property measurements has fueled research into microfluidic rheometry \cite{PM09,GCP07}. Similarly, the characteristics of electrically actuated {microfluidic flows}, specifically measuring the thickness of the Debye layer, depend strongly on non-Newtonian (e.g., shear-thinning or shear-thickening) response of the fluid \cite{ZY13,KPW12}. In applications to lab-on-a-chip technologies, microchannels are used to concentrate or separate particle suspensions, in which case the viscoelastic (non-Newtonian) aspects of the carrier liquid can be employed to achieve this goal \cite{DAGM17}. In the domain of computational methods for non-Newtonian flows, the classical numerical schemes need to be updated to handle the changes in the complex fluid properties using molecular dynamics coupled to continuum models via concurrent simulation or multiscale methods \cite{BLR13}.

An emerging area of {microfluidics} is \emph{fluid--structure interactions} (FSIs). These FSIs occur due to the softness (elastic compliance) of the materials from which flow conduits are manufactured \cite{KCC18}. For example, polydimethylsiloxane (PDMS), a polymeric gel widely used in microfabrication \cite{MW02}, can have a Young's modulus of elasticity on the order of only $1$ MPa \cite{LSC09} (depending on the manufacturing process and conditions), compared to hundreds of GPa for structural steel. Similarly, elastin, a major constituent of blood vessels, is also highly compliant with a Young's modulus $<1$ MPa \cite{SGF77}. FSIs affect fluid flow within a compliant conduit; specifically, the flow rate--pressure drop relation becomes nonlinear \cite{GEGJ06,CCSS17}, and it cannot be described by the classical Hagen--Poiseuille law for rigid conduits \cite{SS93}. When non-Newtonian flows interact with immersed flexible structures, they can cause flutter (a ``side effect'' of FSI that is well known in aerodynamics \cite{BAH96}) even at \emph{vanishing} Reynolds number due to purely elastic flow instabilities \cite{DMSR17}. Further nontrivial interplays occur between electro-osmotic {microfluidic} flows and FSIs: electrokinetics brings about a ``preferential asymmetry to the nature of the dynamical evolution of the relaxation characteristics'' of a deformed microchannel \cite{MCC13}. Such novel physics emerging from fully-coupled flow and deformation must be accurately modeled and incorporated in the design and manufacture of {microfluidic} instruments \cite{KRO14}, such as microvalves \cite{YBP99} and non-invasive pressure sensors \cite{OYE13}.

Recently, Raj and Sen \cite{RS16} documented an initial foray into the experimental interrogation of non-Newtonian FSIs in a microchannel. A rectangular microchannel with three rigid walls and a compliant top wall was manufactured from polydimethylsiloxane (PDMS). A $0.1 \%$ polyoxyethylene (PEO) solution was used as a shear-thinning  non-Newtonian fluid. The pressure drop across $12$ mm increments of the $30$ mm-long microchannel was measured by a series of differential pressure sensors. (The microchannel was long and thin, having a width between $0.5$ and $2$ mm and height of $83$ $\mu$m in the cross-section.) The deflection of the compliant top wall was measured using fluorescence imaging. Raj and Sen \cite{RS16} then proposed a modeling approach for capturing the pressure drop as a function of the flow rate (and other material and geometric parameters) based on a modification of the early work by Gervais \textit{et al.}~\cite{GEGJ06}. The original approach in \cite{GEGJ06} is based on scaling arguments and, thus, contains a fitting parameter that cannot be predicted \textit{a priori}. The approach in \cite{RS16} removes this ambiguity but a narrowly-applicable correlation for stretching of thin shells \cite[pp.~29--33]{S11} is used to replace the pressure--deformation scaling relationship from \cite{GEGJ06}. Thus, Raj and Sen \cite{RS16} do not obtain a flow rate--pressure drop prediction of much generality. {Moreover, since the theoretical predictions were for Newtonian fluids only, they could not be compared with the  experiments reported for non--Newtonian fluids.} Thus, a clear gap remains in the non-Newtonian fluid mechanics literature: \emph{What is the flow rate--pressure drop relation when FSI, due to a non-Newtonian fluid flow in a compliant rectangular microchannel, is taken into account?}

A rigorous mathematical model for viscous non-Newtonian FSIs at the {micron scale} came recently as Boyko \textit{et al.}~\cite{BBG17} studied the flow of a power-law fluid within an elastic cylinder. The unsteady flow field was found under the lubrication approximation. The structural deformation was calculated directly from the governing equations of linear elasticity in a quasi-steady manner under the assumptions of the Love--Kirchhoff hypothesis in classical shell theory. 
An unsteady $p$-Laplacian equation governing the pressure during non-Newtonian FSI was obtained, and its solutions were analyzed in detail. However, the discussion in \cite{BBG17} did not include any experimental or computational verification (benchmarking) of the theoretical predictions. The case of  rectangular microchannels was also not addressed.

Most recently, Kiran Raj \textit{et al.}~\cite{RCDC18} carried out another experimental study of non-Newtonian FSIs. They used 0.04\% by weight solution of Xanthan gum into deionized water as a blood-analog fluid with shear-thinning properties. A 27 mm-long microtube of diameter $\approx 500$ $\mu$m was fabricated from PDMS using a pull-out soft lithography technique. A theory to calculate the pressure drop, as a function of the imposed flow rate in the deformable microtube, was also proposed based on the steady equations of a power-fluid and an elastic half-space. However, the flow regimes investigated in \cite{RCDC18} exhibited only weak FSI, and thus the deviations from the ideal Hagen--Poiseuille law were small. Moreover, the theory proposed in \cite{RCDC18} applies to ``one-way'' coupling only: that is to say, the  pressure was not determined from the flow rate but instead it was measured experimentally, from which the deformation could, ultimately, be obtained. 

Clearly, {microfluidic} FSIs caused by the internal flow of a complex fluid is a research field still in its infancy. More importantly,  previous studies of FSIs due to non-Newtonian fluids  \cite{BBG17,RCDC18} were concerned with (cylindrical) microtubes. In the microfluidics industry, microchannels of (undeformed) rectangular cross-section are more common because of their ease of fabrication \cite{C10_book}. We aim to address the apparent ``knowledge gap'' in {existing research literature} regarding the modeling of non-Newtonian FSIs in rectangular microchannels. 

To this end, in this paper, we develop the theory for steady FSI between a non-Newtonian fluid and the soft elastic top wall of a long, shallow microchannel. The non-Newtonian fluid is assumed to obey a power-law rheological model. The geometry of the microchannel is taken to mimic typical shapes encountered in microfluidics experiments and modeling \cite{GEGJ06,RS16,CCSS17,SC18}, with one compliant wall (top) and three rigid walls (sides and bottom). Specifically this paper is organized as follows. In Sec.~\ref{sec:flow}, the fluid flow problem is solved under the lubrication approximation. Next, following the perturbative approach laid out in \cite{CCSS17}, the thin and thick plate theories of plate bending are invoked to calculate the deflection of the microchannel's top wall due to the hydrodynamic pressure (Sec.~\ref{sec:structural}). We then use these expressions to determine the pressure gradient across the microchannel as a function of the flow rate in Sec.~\ref{sec:finding_q_dp}. However, the pressure itself cannot be explicitly obtained in closed form, \emph{unlike} the case of Newtonian fluids \cite{CCSS17,SC18}. Finally, specializing to two model microchannel geometries, examples of and comparisons between theory and numerical simulation are presented in Sec.~\ref{sec:Results} for thin top walls (Sec.~\ref{sec:thin}) and thick top walls (Sec.~\ref{sec:thick}). In particular, we show that both the flow and structural response results from our perturbative theory compare favorably with full-fidelity, three-dimensional direct numerical simulations of non-Newtonian FSIs.

\section{Problem formulation and mathematical analysis}

\subsection{Fluid flow problem}
\label{sec:flow}

\begin{figure}
  \centering
  \includegraphics[width=0.9\linewidth]{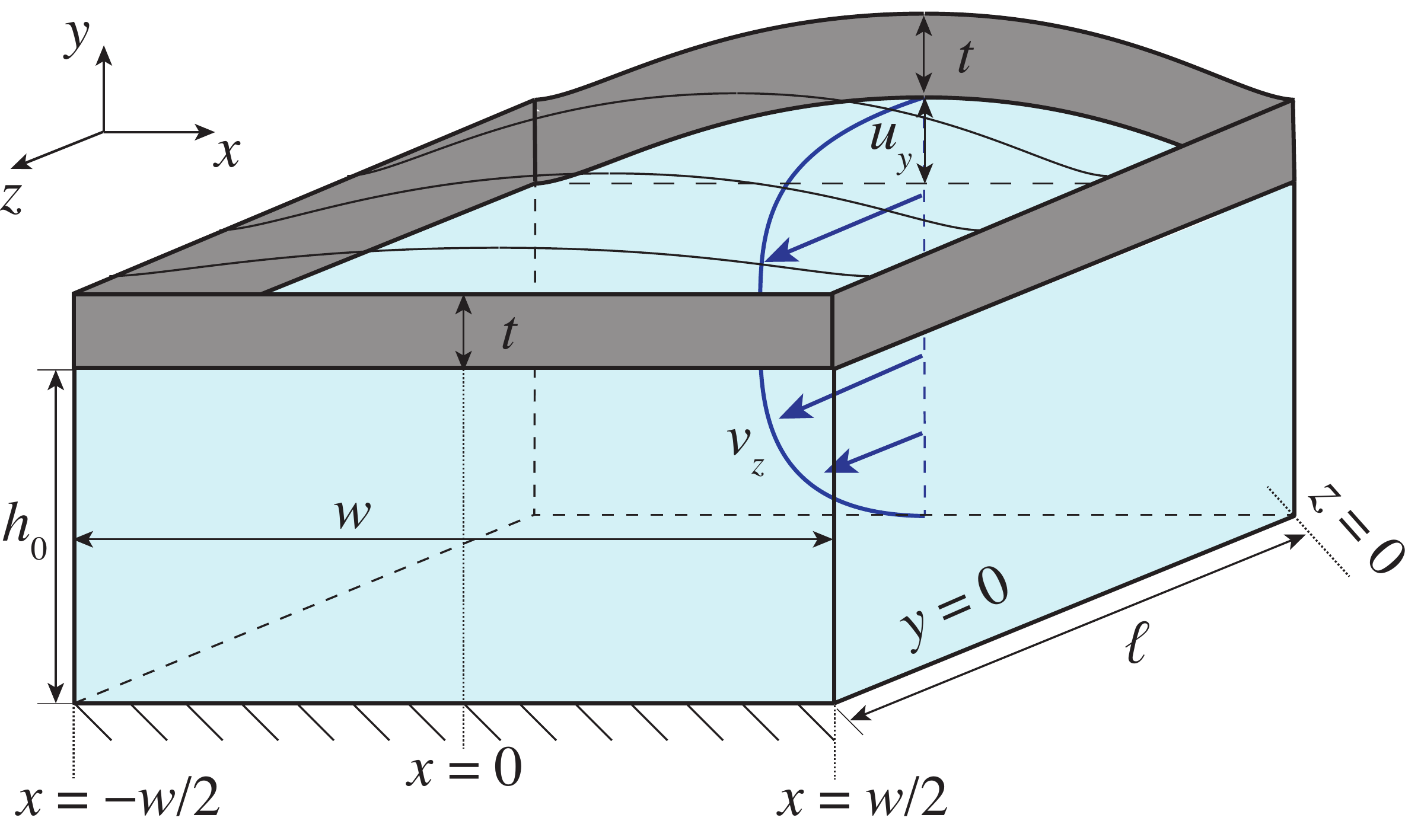}
\caption{Schematic of the chosen microchannel geometry with three rigid and one soft (top) wall. Notation used in the main text is labeled in the schematic. Clamping of the plate along the inlet ($z=0$) plane is not shown for clarity.}
\label{fig:schematic}
\end{figure}

The key assumptions imposed on the fluid flow problem are that an incompressible viscous fluid flows in a channel that is (a) shallow {(width $w$ much greater than its undeformed height $h_0$)} and (b) long {(undeformed height $h_0$ much smaller than its length $\ell$)}; see Fig.~\ref{fig:schematic} for notation. These assumptions lead to an ordering of length scales:
\begin{equation}
h_0\ll w\ll \ell,
\end{equation}
and the emergence of \emph{two} small dimensionless parameters in the problem, i.e., $\epsilon:={h_0}/{\ell}$ and $\delta:={h_0}/{w}$. 
Then, as shown in \cite{CCSS17}, for steady, fully developed conditions, the flow obeys the classical lubrication approximation \cite{panton,B08} for $\epsilon\ll\delta\ll1$. Specifically, to the leading order in $\epsilon$ and $\delta$, the flow field is unidirectional 
\begin{equation}
\label{velocity_mt_1}
   \bm{v} \sim v_z(x,y) \bm{\hat{k}},
\end{equation}
and the $\bm{\hat{\imath}}$ and $\bm{\hat{\jmath}}$ flow components can be thus neglected. (Here, we use the standard notation of $\bm{\hat{\imath}}$, $\bm{\hat{\jmath}}$ and $\bm{\hat{k}}$ being the unit normal vectors in the $x$, $y$ and $z$ coordinate directions.) Such a flow automatically satisfies the conservation of mass (continuity) equation $\bm{\nabla}\bm{\cdot}\bm{v}=0$ for an incompressible fluid.\footnote{This statement is true subject to the caveat that, due to FSI, $v_z$ will ``pick up'' a weak dependence on $z$, which is still a valid unidirectional flow within the lubrication approximation \cite{panton,B08}.} 
Similarly, the lubrication approximation dictates that we may neglect  inertial forces in comparison to viscous and pressure forces in the fluid. Then, as a consequence of the kinematics, i.e., Eq.~\eqref{velocity_mt_1}, the pressure becomes independent of $x$ and $y$, leading to $\rd p/\rd z=const.$ This result is a well known consequence of the flow geometry being ``long and thin'' \cite{D17b}. Lubrication flows of non-Newtonian fluids have been studied extensively, for example in the context of \emph{rivulets} \cite{WDH02,AMWD15}, but not as much in the microfluidics and FSI contexts.

Next, the fluid's only nontrivial momentum balance is in the $z$-direction:
\begin{equation}
0 = - \frac{\rd p}{\rd z} + \frac{\partial \tau_{yz}}{\partial y} \qquad\Rightarrow\qquad \tau_{yz} = \frac{\rd p}{\rd z}y+C_1.
\label{eq:momentum}
\end{equation}
To fix the constant of integration, we must specify $\tau_{yx}$ somewhere. Consider the following: for a rigid channel the centerline is at $y=h_0/2$, however, after the deformation of the upper wall, the centerline shifts to $y=(h_0+u_y)/2$. Here, $u_y(x,z)$ is the absolute deformation of the top wall in the $y$-direction, while $h_0=const.$ is the undeformed (initial) height of the channel (see Fig.~\ref{fig:schematic}). We require that the shear stress vanish at the centerline, $\tau_{yz}|_{y=(h_0+u_y)/2} = 0$. Hence, 
\begin{equation}
C_1=-\frac{\rd p}{\rd z}\left(\frac{h_0+u_y}{2}\right).
\end{equation}

Next, we must specify the constitutive equation between stress and shear rate-of-strain. The rheological behavior of most biofluids is non-Newtonian \cite{Lee06}. Specifically, human blood has been shown to exhibit a shear-thinning behavior \cite{Chien66}, which can be described by the \emph{power-law} (also known as Ostwald--de Waele) model \cite{Bird76}. Under the lubrication approximation \cite{panton,B08}, it can be shown that the dominant term in the deviatoric stress tensor is the shear stress. Under the latter approximation, the power-law rheological model reduces to:
\begin{equation}
\tau_{zy} = \eta\frac{\partial v_z}{\partial y} = m\left|\frac{\partial v_z}{\partial y}\right|^{n-1}\left(\frac{\partial v_z}{\partial y}\right),
\label{eq:constitutive}
\end{equation}
where $\eta = m|{\partial v_z}/{\partial y}|^{n-1} $ is the \emph{apparent viscosity} \cite{BAH87}. Here $m$ is the consistency factor, which reduces to the dynamic viscosity for a Newtonian fluid, and $n$ is the power-law index with a value of $n=1$ for the special case of a Newtonian fluid. 

Now, for the lower half of the channel, i.e., for $0 \le y \le (h_0+u_y)/2 $, we expect the velocity gradient to be positive under our chosen sign convention:
\begin{equation}
\frac{\partial v_z}{\partial y}>0 \qquad\Rightarrow\qquad \left|\frac{\partial v_z}{\partial y}\right|=\frac{\partial v_z}{\partial y}.
\end{equation}
Then, combining the momentum Eq.~\eqref{eq:momentum} with the appropriate simplification of the constitutive Eq.~\eqref{eq:constitutive} yields
\begin{multline}
m\left(\frac{\partial v_z}{\partial y}\right)^n = \left(\frac{\rd p}{\rd z}\right)\left(y-\frac{h_0+u_y}{2}\right)\\
\qquad\Rightarrow\qquad
\frac{\partial v_z}{\partial y} = \left(-\frac{1}{m}\frac{\rd p}{\rd z}\right)\left(\frac{h_0+u_y}{2}-y\right)^{1/n}.
\end{multline}
Integration of the last equation subject to no slip on the lower wall, i.e., $v_z=0$ at $y=0$, yields the expression for velocity profile in the \emph{lower} half of the channel, i.e., $0 \le y \le (h_0+u_y)/2$:
\begin{multline}
v_z(x,y,z) = \frac{1}{1+1/n}\left(-\frac{1}{m}\frac{\rd p}{\rd z}\right)^{1/n}\left(\frac{h_0+u_y}{2}\right)^{1+1/n}\\
\times\left\{1-\left[1-\frac{y}{(h_0+u_y)/2}\right]^{1+1/n}\right\}.
\label{eq:vz_lower}
\end{multline}

Now, for the upper half of the channel, i.e., for $(h_0+u_y)/2 \le y \le (h_0+u_y)$, the velocity gradient is negative:
\begin{equation}
\frac{\partial v_z}{\partial y}<0 \qquad\Rightarrow\qquad \left|\frac{\partial v_z}{\partial y}\right|=-\frac{\partial v_z}{\partial y}.
\end{equation}
Consequently,
\begin{multline}
-m\left(-\frac{\partial v_z}{\partial y}\right)^n = \frac{\rd p}{\rd z}\left(y-\frac{h_0+u_y}{2}\right)\\
\qquad\Rightarrow\qquad
-\frac{\partial v_z}{\partial y}=\left(-\frac{1}{m}\frac{\rd p}{\rd z}\right)^{1/n}\left(y-\frac{h_0+u_y}{2}\right)^{1/n}.
\end{multline}
Integration of the last equation subject to no slip at the upper wall, i.e., $v_z=0$ at $y = h_0+u_y$ yields the expression for velocity profile in the \emph{upper} half of the channel, i.e., $(h_0+u_y)/2 \le y \le h_0+u_y$:
\begin{multline}
v_z(x,y,z) = \frac{1}{1+1/n}\left(-\frac{1}{m}\frac{\rd p}{\rd z}\right)^{1/n}\left(\frac{h_0+u_y}{2}\right)^{1+1/n}\\
\times \left\{1-\left[\frac{y}{(h_0+u_y)/2}-1\right]^{1+1/n}\right\}.
\label{eq:vz_upper}
\end{multline}
This completes the derivation of the velocity profile, which is piecewise defined by Eqs.~\eqref{eq:vz_lower} and \eqref{eq:vz_upper}, of a power-law fluid in a microchannel with a deformed top wall shape given by 
\begin{equation}
h(x,z) = h_0 + u_y(x,z).
\end{equation}

\subsubsection{Nondimensionalization}
To make the governing equations dimensionless, we introduce the following dimensionless variables:
\begin{multline}
\bar{p} = p/\mathcal{P}_c, \quad
\bar{v}_{\bar{z}} = v_z/\mathcal{V}_c, \quad
\bar{x} = x/w, \quad
\bar{y} = y/h_0, \quad
\bar{z} = z/\ell, \\
\bar{u}_{\bar{y}} = u_y/u_c,\quad 
\beta = u_c/h_0, \quad
\epsilon = h_0/\ell, \quad
\delta = h_0/w. \quad
\label{eq:ndvars}
\end{multline}
Here, $h_0$, $w$ and $\ell$ are, respectively, the undeformed height, fixed width and constant length of the microchannel as in Fig.~\ref{fig:schematic}. {Making the fluid's momentum equation~\eqref{eq:momentum} dimensionless and ensuring all terms are of order one, we can determine how the characteristic velocity scale is related to the characteristic pressure scale, namely 
$\mathcal{V}_c = [{\mathcal{P}_c h_0}/(m\ell)]^{1/n} h_0$. Then, the pressure scale can be taken to be $\mathcal{P}_c = \Delta p$ if $\Delta p$ is known (pressure-controlled experiment/simulation) or estimated as $\mathcal{P}_c = [q/(h_0^2w)]^n m \ell/h_0$ if $q$ is known (flow-rate-controlled experiment/simulation). Each choice then yields a corresponding velocity scale $\mathcal{V}_c=[{\Delta p\, h_0}/(m\ell)]^{1/n} h_0$ or $\mathcal{V}_c=q/(h_0w)$, respectively.} The characteristic scale of deflection $u_c$ will be determined self-consistently below by coupling the flow and structural deformation problems, from which we will then determine the ratio $\beta$.  The two length ratios, $\epsilon$ and $\delta$, are key to satisfying the asymptotic requirements of the lubrication approximation. Specifically, we require that a separation of scales exists, i.e., $0< \epsilon \ll \delta \ll 1$, meaning the microchannel is \emph{long} and \emph{shallow}. 

In terms of the dimensionless variables from Eq.~\eqref{eq:ndvars}, the velocity profile from Eqs.~\eqref{eq:vz_lower} and \eqref{eq:vz_upper} above now takes the form:
\begin{multline}
\label{velocity_non_dim}
\bar{v}_{\bar{z}}(\bar{x},\bar{y},\bar{z}) = 
\frac{1}{2^{1+1/n}}\left(-\frac{\rd\bar{p}}{\rd\bar{z}}\right)^{1/n}\frac{(1+\beta\bar{u}_{\bar{y}})^{1+1/n}}{1+1/n}\\
\times
\begin{cases}
\displaystyle  1 - \left[1-\frac{\bar{y}}{(1+\beta\bar{u}_{\bar{y}})/2}\right]^{1+1/n}, & \bar{y} <[1+\beta\bar{u}_{\bar{y}}(\bar{x},\bar{z})]/2,\\[4mm]
\displaystyle 1-\left[\frac{\bar{y}}{(1+\beta\bar{u}_{\bar{y}})/2}-1\right]^{1+1/n}, & \bar{y} \ge [1+\beta\bar{u}_{\bar{y}}(\bar{x},\bar{z})]/2,
\end{cases}
\end{multline}
while the dimensionless microchannel height is 
\begin{equation}
\bar{h}(\bar{x},\bar{z}) \equiv {h(x,z)}/{h_0} = 1 + \beta \bar{u}_{\bar{y}}(\bar{x},\bar{z}).
\label{eq:h_nondim}
\end{equation}
It is straightforward to verify that the velocity profile is continuous at $\bar{y}=(1+\beta\bar{u}_{\bar{y}})/2$.

\subsection{Coupling flow and deformation: Finding the flow rate} 
We study steady flow, thus conservation of mass then requires that the volumetric flow rate $q=const.$ This flow rate is, by definition, given by
\begin{equation}\label{eq:q1}
\begin{aligned}
q &= \int\limits_{-w/2}^{+w/2} \int\limits_{0}^{h_0+u_y} v_z(x,y,z) \,\rd x\, \rd y \\
&= \int\limits_{-w/2}^{+w/2} \int\limits_{0}^{(h_0+u_y)/2} v_z \,\rd x\, \rd y + \int\limits_{-w/2}^{+w/2} \int\limits_{(h_0+u_y)/2}^{h_0+u_y} v_z \,\rd x \, \rd y,
\end{aligned}
\end{equation}
where we have split the integral over $y$ to account for the velocity profile's slope change at $y=(h_0+u_y)/2$. A switch to dimensionless variables turns Eq.~\eqref{eq:q1} into
\begin{equation}\label{eq:q2}
q = \mathcal{V}_c h_0 w \left\{\, \int\limits_{-1/2}^{+1/2} \int\limits_{0}^{(1+\beta\bar{u}_{\bar{y}})/2}\bar{v}_{\bar{z}} \,\rd\bar{x} \,\rd\bar{y} + \int\limits_{-1/2}^{+1/2} \int\limits_{(1+\beta\bar{u}_{\bar{y}})/2}^{1+\beta\bar{u}_{\bar{y}}}\bar{v}_{\bar{z}} \,\rd\bar{x} \,\rd\bar{y}\right\}.
\end{equation}
Substituting the expressions for $\bar{v}_{\bar{z}}$ from Eq.~\eqref{velocity_non_dim} into Eq.~\eqref{eq:q2} and re-arranging:
\begin{multline}\label{eq:q3}
\frac{q}{\mathcal{V}_c h_0 w} = \frac{1}{2^{1+1/n}(1+1/n)} \left(-\frac{\rd\bar{p}}{\rd\bar{z}}\right)^{1/n} \int\limits_{-1/2}^{+1/2} (1+\beta\bar{u}_{\bar{y}})^{1+1/n} \\
\times \left\{\int\limits_{0}^{(1+\beta\bar{u}_{\bar{y}})/2}\left(1-\left[1-\frac{\bar{y}}{(1+\beta\bar{u}_{\bar{y}})/2}\right]^{1+1/n}\right)\rd\bar{y} \right. \\
\left. + \int\limits_{(1+\beta\bar{u}_{\bar{y}})/2}^{1+\beta\bar{u}_{\bar{y}}} \left(1-\left[\frac{\bar{y}}{(1+\beta\bar{u}_{\bar{y}})/2}-1\right]^{1+1/n}\right) \rd\bar{y}\right\} \rd\bar{x}.
\end{multline}
Let us now introduce the dimensionless flow rate $\bar{q} = q/(\mathcal{V}_c h_0 w)$.\footnote{Of course, since $\mathcal{V}_c=q/(h_0w)$ for a flow-rate-controlled experiment or simulation, this definition of $\bar{q}$ would yield $\bar{q}=1$ in this case. Nevertheless, we keep $\bar{q}$ in the equations that follow to allow the reader to easily use them under a pressure-drop-controlled situation, or with a different $\mathcal{V}_c$ expression.} Then, performing the integration in Eq.~\eqref{eq:q3} along with the requisite substitution of limits yields
\begin{equation}
\bar{q} = \frac{1}{2^{1+1/n}(2+1/n)}\left(-\frac{\rd\bar{p}}{\rd\bar{z}}\right)^{1/n}\int\limits_{-1/2}^{+1/2}(1+\beta\bar{u}_{\bar{y}})^{2+1/n} \,\rd\bar{x}.
\label{eq:channel_Q_w_int}
\end{equation}
To complete the calculation, we must specify the cross-sectional displacement profile $\bar{u}_{\bar{y}}=\bar{u}_{\bar{y}}(\bar{x},\bar{z})$.

\subsection{Structural mechanics problem}
\label{sec:structural}

Typical manufacturing techniques for PDMS-based microfluidics yield microchannels and top walls that are rectangular \cite{BW16}. Thus, assuming isotropic and quasi-static plate-bending under steady fluid flow, the governing equations of the solid mechanics problems can be taken to be those of either the zeroth-order (i.e., Kirchhoff--Love (KL) \cite{L88,TWK59}) or the first-order shear-deformation (i.e., Reissner--Mindlin (RM) \cite{R45,M51}) plate theory. Kirchhoff--Love plate theory, which preserves all the assumptions of Love's first postulate,  is the straightforward extension of Euler beam theory to two spatial dimensions. Thus, KL theory only accounts for plane stress and plane strain condition inside the plate. More specifically, the KL theory applies when the maximum top wall deformation $u_{y,\mathrm{max}}$ is such that $u_{y,\mathrm{max}}\ll t \ll w$, i.e., the KL theory is a \emph{thin-plate theory}. On the other hand, Reissner--Mindlin theory allows for shear deformations (and strains) across a  plate of finite thickness. More specifically, the RM theory applies when $u_{y,\mathrm{max}}\ll t \simeq w$, i.e., the RM theory is a \emph{thick-plate theory}. Both these theories are \emph{linear}, in the sense that strains ($\sim u_{y,\mathrm{max}}/w$) are assumed small, so that the strain--displacement relations are linear, and the constitutive equation relating the stress to the strain is also linear. 

Under the same lubrication scaling ($h_0 \ll w \ll \ell$) as for the fluid mechanics problem, it was shown in \cite{CCSS17,SC18}, that the flow-wise and span-wise displacements \emph{decouple}. Then, to the leading order in $\epsilon$ and $\delta$, the displacement of the top wall is given by the solution of a beam equation in the $({x},{y})$ cross-section (fixed ${z}$) subject to the applied (constant in the cross-section) load ${p}({z})$. In terms of the dimensionless variables from Eq.~\eqref{eq:ndvars}, the  displacement profile is thus given by either
\begin{align}
\label{eq:Deflection_Channel}
\text{KL}:\quad \bar{u}_{\bar{y}}(\bar{x},\bar{z}) &= \frac{1}{24} \left(\frac{1}{4}-\bar{x}^2\right)^2 \bar{p}(\bar{z}),\\
\label{eq:Deflection_Channel2}
\text{RM}:\quad \bar{u}_{\bar{y}}(\bar{x},\bar{z}) &= \frac{1}{24}  \left(\frac{1}{4} - \bar{x}^2\right) \left[ \frac{2(t/w)^2}{\kappa(1-\nu_\mathrm{s})}  + \left(\frac{1}{4} - \bar{x}^2\right) \right] \bar{p}(\bar{z}).
\end{align}
Notice that Eq.~\eqref{eq:Deflection_Channel} can be understood simply as the $t/w\to0^+$ limit of Eq.~\eqref{eq:Deflection_Channel2}, which captures the effect of finite plate thickness $t$. In Eq.~\eqref{eq:Deflection_Channel2}, $\kappa$ is a \emph{shear-deformation factor} \cite{C66,H01}, which must be taken as unity (i.e., $\kappa=1$) based on previous mathematical analysis \cite{Z06} that was further justified in \cite{SC18} through comparisons with simulations and experiments.

From Eq.~\eqref{eq:Deflection_Channel2}, it is clear that $\bar{u}_{\bar{y}}$ increase with $t/w$ (at a \emph{fixed} load $\bar{p}$). {To understand this observation, we recall that RM plate theory allows the normal to the plate's reference surface to change direction (and, thus, no longer remain normal, as it would in the KL theory) during deformation. Shear strains in the transverse (to the surface) direction are thus allowed and accounted for. Then, since shear deformations are integrals of shear strains along the thickness, increasing the thickness leads to larger shear strains and, thereby, larger overall deformation in the normal direction, (see also \cite[Ch.~13]{ZTZ13})}.

An important feature of the asymptotic decoupling introduced in \cite{CCSS17} is that the structural deformation equations, and the resulting deflection profiles, are \emph{not explicitly} dependent on the fluid's rheology! The only fluid quantity that enters Eqs.~\eqref{eq:Deflection_Channel} and \eqref{eq:Deflection_Channel2} is the pressure $\bar{p}$. Of course, the latter is computed from the flow profile, which depends on the rheological model at hand, but none of this information is required to write down the deflection profiles explicitly. Here, we shall not expend any more effort on the structural mechanics problem as it is already developed in full detail in \cite{CCSS17,SC18}.

In both of the analyses \cite{CCSS17,SC18}, the characteristic deformation scale turns out to be $u_c={w^4\mathcal{P}_c}/{B}$, where $B = {Et^3}/[12(1-\nu_\mathrm{s}^2)]$ is the \emph{bending rigidity} of a plate \cite{TWK59}. This choice ensures a leading-order asymptotic balance \cite{CCSS17}. Hence, we can now express $\beta(>0)$ defined in Eq.~\eqref{eq:ndvars} explicitly as
\begin{equation}
\beta := \frac{u_c}{h_0} = \frac{w^4\mathcal{P}_c}{Bh_0} \equiv  \frac{w^{4-n} \ell m q^n}{Bh_0^{2(1+n)}}.
\label{eq:beta}
\end{equation}
Since $\beta$ is a dimensionless ratio of the hydrodynamic forces applied on the solid ($\mathcal{P}_c\propto\Delta p$) to its bending rigidity, we term $\beta$ the \emph{FSI parameter}. It should be clear from Eq.~\eqref{eq:h_nondim} that $\beta = 0$ corresponds to a rigid (undeformable) top wall.

\subsection{Completing the calculation}
\label{sec:finding_q_dp}

Although $\bar{u}_{\bar{y}}$, as given by Eqs.~\eqref{eq:Deflection_Channel} and \eqref{eq:Deflection_Channel2}, is ``simply'' a quartic polynomial in $\bar{x}$, the remaining integral in the flow rate expression from Eq.~\eqref{eq:channel_Q_w_int} cannot be performed in terms of elementary functions for arbitrary $n$ because $2+1/n$ is a \emph{fractional} power, even for integer $n$. On the other hand, a binomial expansion of the integrand yields 
\begin{multline}
(1+\beta\bar{u}_{\bar{y}})^{2+1/n} = \sum_{k=0}^{\infty} C_{k,n}(\beta\bar{u}_{\bar{y}})^k,\\ C_{k,n} := \frac{\Gamma (3+1/n)}{\Gamma (k+1) \Gamma (1/n-k+3)},
\label{eq:integrand_binomial}
\end{multline}
where $C_{k,n}$ is the binomial coefficient \cite{W18_BC}, and $\Gamma(\zeta) := \int_0^\infty s^{\zeta-1}\re^{-s}\,\rd s$ is the Gamma function. Clearly, if $2+1/n$ is an integer, then the expansion must terminate for $k-1$ such that $1/n-k+3=0$; for example, it terminates at $k=3$ for $n=1$. 

For non-integer $2+1/n$, the expansion in Eq.~\eqref{eq:integrand_binomial} is exact if taken as a convergent infinite sum, { which is the case only if $\beta \bar{u}_{\bar{y}} < 1$. We expect (and the results below confirm) that for the typical displacements and values of $\beta$ relevant to microchannel FSI, the convergence criterion is satisfied.} Finally, in practice, the infinite sum in Eq.~\eqref{eq:integrand_binomial} must be truncated. We now undertake the endeavor of determining how and when this truncation can be done.

\subsubsection{Negligible plate thickness ($t/w\to0$)}
\label{sec:integration1}

Although it is possible to unify the cases of thin and thick plates from the start, by using Eq.~\eqref{eq:Deflection_Channel} for the displacement, it is quite illustrative to perform the thin and thick plate calculations separated, due to the simplifications arising in the integration of various terms. To this end, substituting the expression for $\bar{u}_{\bar{y}}$ from Eq.~\eqref{eq:Deflection_Channel} into Eq.~\eqref{eq:integrand_binomial} yields
\begin{equation}
(1+\beta\bar{u}_{\bar{y}})^{2+1/n} =\sum_{k=0}^{\infty}C_{k,n}\left(\frac{\beta \bar{p}}{24}\right)^k(\bar{x}+1/2)^{2k}(\bar{x}-1/2)^{2k}.
\end{equation}
Now, let us consider just the integral
\begin{equation}
\begin{aligned}
&\int\limits_{-1/2}^{+1/2} (1+\beta\bar{u}_{\bar{y}})^{2+1/n} \,\rd\bar{x} \\
&= 
\int\limits_{-1/2}^{+1/2} \sum_{k=0}^{\infty}C_{k,n} \left(\frac{\beta \bar{p}}{24}\right)^k(\bar{x}+1/2)^{2k}(\bar{x}-1/2)^{2k} \,\rd\bar{x} \\
&= \sum_{k=0}^{\infty}C_{k,n} \left(\frac{\beta \bar{p}}{24}\right)^k \int\limits_{-1/2}^{+1/2} (\bar{x}+1/2)^{2k}(\bar{x}-1/2)^{2k} \,\rd\bar{x}\\
&= \sum_{k=0}^{\infty}C_{k,n} \left(\frac{\beta \bar{p}}{24}\right)^k \frac{k \sqrt{\pi}\, \Gamma(2 k)}{2^{4 k} \,\Gamma(3/2 + 2 k)}.
\end{aligned}
\end{equation}
Then, Eq.~\eqref{eq:channel_Q_w_int} becomes
\begin{multline}
\label{Pressure_Drop_Vs_FlowRate_Microchannel}
\bar{q} = \frac{1}{2^{1+1/n}(2+1/n)}\left(-\frac{\rd\bar{p}}{\rd\bar{z}}\right)^{1/n} \\
\times \sum_{k=0}^{\infty}C_{k,n} \left(\frac{\beta \bar{p}}{24}\right)^k \frac{k \sqrt{\pi}\, \Gamma(2 k)}{2^{4 k} \,\Gamma(3/2 + 2 k)}.
\end{multline}
Finally, we re-express the latter equation as  a first-order ODE:
\begin{multline}
\frac{\rd\bar{p}}{\rd\bar{z}} = -\bar{q}^n \left\{ \frac{1}{2^{1+1/n}(2+1/n)} \right. \\
\times \left. \sum_{k=0}^{\infty}C_{k,n} \left[\frac{\beta}{24}\bar{p}(\bar{z})\right]^k \frac{k \sqrt{\pi}\, \Gamma(2 k)}{2^{4 k} \,\Gamma(3/2 + 2 k)}\right\}^{-n}.
\label{eq:p_ode_thin_plate}
\end{multline}
All quantities, except $\bar{p}$, on the right-hand side of the latter equation are given constants, thus Eq.~\eqref{eq:p_ode_thin_plate} is indeed a first-order \emph{nonlinear} ODE. 

The ODE~\eqref{eq:p_ode_thin_plate} is obviously separable but the integration cannot be carried out, even with special functions, due to the infinite sum over $k$. Instead, below we will integrate the ODE~\eqref{eq:p_ode_thin_plate} numerically, checking to see how many terms in the $k$-summation are needed to obtain results insensitive to the truncation of the infinite sum.

\subsubsection{Consistency checks}
\label{sec:checks1}

Although verifying that a mathematical expression reduces to a previous result in a special case cannot speak to the veracity of said mathematical expression, considering some limiting cases of our theory is of pedagogical value.

First, the rigid-channel limit is easy to take from Eq.~\eqref{Pressure_Drop_Vs_FlowRate_Microchannel} by letting $\beta \to 0^+$:
\begin{equation}
\bar{q} = \frac{1}{2^{1+1/n}(2+1/n)}\left(-\frac{\rd\bar{p}}{\rd\bar{z}}\right)^{1/n}.
\label{eq:q_dpdz_beta0}
\end{equation}
This flow rate--pressure gradient expression for a power-law fluid in a slot can be matched to the one given in the first row of Table 4.2-1 in \cite{BAH87}.

Second, the Newtonian limit is also easy to take, starting from Eq.~\eqref{Pressure_Drop_Vs_FlowRate_Microchannel} and noting that the sum terminates at $k=3$ when $n=1$, to obtain:
\begin{equation}
\bar{q} = - \frac{1}{12}\frac{\rd\bar{p}}{\rd\bar{z}} \left[\sum_{k=0}^{3} C_{k,1} \left(\frac{\beta \bar{p}}{24}\right)^k \frac{k \sqrt{\pi}\, \Gamma(2 k)}{2^{4 k} \,\Gamma(3/2 + 2 k)}\right].
\label{eq:channel_Q_n1}
\end{equation}
Subjecting Eq.~\eqref{eq:channel_Q_n1} to the outlet boundary condition $\bar{p}(1) = 0$, it can be separated and integrated to yield:
\begin{equation}
\bar{q} =  \frac{1}{12}\frac{\bar{p}(\bar{z})}{(1-\bar{z})} \left\{\sum_{k=0}^{3}\frac{C_{k,1}}{k+1} \left[\frac{\beta}{24}\bar{p}(\bar{z})\right]^k \frac{k \sqrt{\pi}\, \Gamma(2 k)}{2^{4 k} \,\Gamma(3/2 + 2 k)}\right\}.
\label{eq:channel_Q_n1_integrated}
\end{equation}
The {implicit algebraic} relation for $\bar{p}(\bar{z})$ in Eq.~\eqref{eq:channel_Q_n1_integrated} can be compared to \cite[Eq.~(4.3)]{CCSS17}, matching the coefficients of $\bar{p}^k$, $k=0,1,2,3$ (see \ref{app:coeff_match} for more details).

Third, letting $\beta\to0^+$ in Eq.~\eqref{eq:channel_Q_n1_integrated} means that only the $k=0$ term in the summation ``survives,'' and we obtain
\begin{equation}
\bar{q} = \frac{1}{12}\frac{\bar{p}(\bar{z})}{(1-\bar{z})}  \lim_{k\to 0} \left[ \frac{C_{k,1}}{k+1} \frac{k \sqrt{\pi}\, \Gamma(2 k)}{2^{4 k} \,\Gamma(3/2 + 2 k)}\right].
\end{equation}
The limit on the right-hand side can be shown to equal 1. Hence, we have recovered the pressure profile for unidirectional flow in a slot of height $h_0$ and width $w$: $\bar{p}(\bar{z}) = 12\bar{q}(1-\bar{z})$. The latter, evaluated at $\bar{z}=0$ along with the definition of $\mathcal{V}_c$ substituted back in, yields
$\Delta p = 12\mu \ell q/(h_0^3 w)$, where $\Delta p := p(0) - p(1)$ is the full pressure drop across the microchannel, as expected \cite[\S3.4.2]{B08}.

\subsubsection{Plates with considerable thickness (finite $t/w$)}
\label{sec:integration2}

As in Sec.~\ref{sec:integration1}, let us consider just the integral in Eq.~\eqref{eq:channel_Q_w_int} but with the top-wall deformation from Eq.~\eqref{eq:Deflection_Channel2} under the binomial series expansion from Eq.~\eqref{eq:integrand_binomial}:
\begin{multline}
\int\limits_{-1/2}^{+1/2} (1+\beta\bar{u}_{\bar{y}})^{2+1/n} \,\rd\bar{x} \\ 
= \sum_{k=0}^{\infty}C_{k,n} \left(\frac{\beta \bar{p}}{24}\right)^k \int\limits_{-1/2}^{+1/2} (1/4-\bar{x}^2)^{k} \left[ \frac{2(t/w)^2}{\kappa(1-\nu_\mathrm{s})}  + \left(\frac{1}{4} - \bar{x}^2\right) \right]^{k} \,\rd\bar{x}\\
= \sum_{k=0}^{\infty}C_{k,n} \left(\frac{\beta \bar{p}}{24}\right)^k2^{-1-4k}\left(1+\frac{8(t/w)^2}{\kappa(1-\nu_\mathrm{s})}\right)^{k}\sqrt{\pi}\, \Gamma{(k+1)}\\
\times \,{}_2\tilde{F}_1\left(\frac{1}{2},-k;\frac{3}{2}+k;\frac{1}{1+\frac{8(t/w)^2}{\kappa(1-\nu_\mathrm{s})}}\right),
\end{multline}
\\
where ${}_2\tilde{F}_1$ is the \emph{regularized hypergeometric function} \cite{W18_RHF}. 
Therefore, the flow rate becomes
\begin{multline}
\label{eq:q_thick_plate}
\bar{q} = \frac{1}{2^{1+1/n}(2+1/n)}\left(-\frac{\rd\bar{p}}{\rd\bar{z}}\right)^{1/n}
\left\{\sum_{k=0}^{\infty}C_{k,n} \left(\frac{\beta \bar{p}}{24}\right)^k2^{-1-4k} \right.\\
\times \left[1+\frac{8(t/w)^2}{\kappa(1-\nu_\mathrm{s})}\right]^{k}\sqrt{\pi} \, \Gamma{(k+1)}\\
\left. \times \,{}_2\tilde{F}_1\left(\frac{1}{2},-k;\frac{3}{2}+k;\frac{1}{1+\frac{8(t/w)^2}{\kappa(1-\nu_\mathrm{s})}}\right)\right\}.
\end{multline}
Again, Eq.~\eqref{eq:q_thick_plate} can be put in the form of an ODE for the pressure:
\begin{multline}
\frac{\rd\bar{p}}{\rd\bar{z}} = -\bar{q}^n \left\{\frac{1}{2^{1+1/n}(2+1/n)} \sum_{k=0}^{\infty}C_{k,n} \left[\frac{\beta}{24}\bar{p}(\bar{z})\right]^k2^{-1-4k} \right.\\
\times \left[1+\frac{8(t/w)^2}{\kappa(1-\nu_\mathrm{s})}\right]^{k} \sqrt{\pi} \, \Gamma{(k+1)}\\
\left. \times \,{}_2\tilde{F}_1\left(\frac{1}{2},-k;\frac{3}{2}+k;\frac{1}{1+\frac{8(t/w)^2}{\kappa(1-\nu_\mathrm{s})}}\right)\right\}^{-n}.
\label{eq:p_ode_thick_plate}
\end{multline}

\subsubsection{Consistency checks}
\label{sec:checks2}

First, the rigid-channel limit is obtained exactly as in Sec.~\ref{sec:checks1} to yield Eq.~\eqref{eq:q_dpdz_beta0}.

Second, the Newtonian limit is easy to take starting from Eq.~\eqref{eq:q_thick_plate}, noting that the sum terminates at $k=3$ when $n=1$:
\begin{multline}
\bar{q} = - \frac{1}{12}\frac{\rd\bar{p}}{\rd\bar{z}} \left\{\sum_{k=0}^{3} C_{k,1} \left(\frac{\beta \bar{p}}{24}\right)^k2^{-1-4k} \left[1+\frac{8(t/w)^2}{\kappa(1-\nu_\mathrm{s})}\right]^{k} \right.\\
 \left. \times \sqrt{\pi}\, \Gamma{(k+1)} \,{}_2\tilde{F}_1\left(\frac{1}{2},-k;\frac{3}{2}+k;\frac{1}{1+\frac{8(t/w)^2}{\kappa(1-\nu_\mathrm{s})}}\right) \right\}.
\label{eq:q_dpdz_thick_Newtonian}
\end{multline}
For a pressure distribution subject to the outlet boundary condition $\bar{p}(1)=0$, Eq.~\eqref{eq:q_dpdz_thick_Newtonian} can be separated and integrated to yield:
\begin{multline}
\bar{q} =  \frac{1}{12}\frac{\bar{p}(\bar{z})}{(1-\bar{z})} \left\{\sum_{k=0}^{3}\frac{C_{k,1}}{k+1} \left[\frac{\beta}{24}\bar{p}(\bar{z})\right]^k2^{-1-4k} \left[1+\frac{8(t/w)^2}{\kappa(1-\nu_\mathrm{s})}\right]^{k} \right.\\
\left. \times \sqrt{\pi}\, \Gamma{(k+1)} \,{}_2\tilde{F}_1\left(\frac{1}{2},-k;\frac{3}{2}+k;\frac{1}{1+\frac{8(t/w)^2}{\kappa(1-\nu_\mathrm{s})}}\right) \right\}.
\label{eq:channel_Q_thick_n1}
\end{multline}
Equation~\eqref{eq:channel_Q_thick_n1} is an implicit algebraic equation for $\bar{p}(\bar{z})$. It can be compared to \cite[Eqs.~(17)--(20)]{SC18}, matching the coefficients of $\bar{p}^k$, $k=0,1,2,3$ (see \ref{app:coeff_match} for more details).

Third, the thickness contribution can be neglected by setting $t/w \equiv 0$. Then, Eq.~\eqref{eq:q_thick_plate} reduces to Eq.~\eqref{eq:p_ode_thin_plate}, derived for a thin-plate top wall, upon taking into account the appropriate limiting behaviors of ${}_{2}\tilde{F}_1$ \cite{W18_RHF}.

\section{Results and discussion}
\label{sec:Results}

\subsection{Illustrative examples of the theoretical results}
\label{sec:theory_examples}

As mentioned in Sec.~\ref{sec:intro}, the ultimate objective of the present work is to investigate steady-state FSIs due to {internal flow of a power-law fluid} within a microchannel with a soft top wall. In this subsection, we discuss and illustrate the interplay between the FSI parameter $\beta$ [recall Eq.~\eqref{eq:beta}], the fluid's power-law index $n$,  and the thickness-to-width ratio $t/w$ of the top wall. Specifically, we address how the latter interplay sets the dimensionless pressure drop $\Delta \bar{p}$ over the microchannel for a given dimensionless flow rate $\bar{q}$. 

\begin{figure}[ht!]
	\centering
	\includegraphics[width=\linewidth]{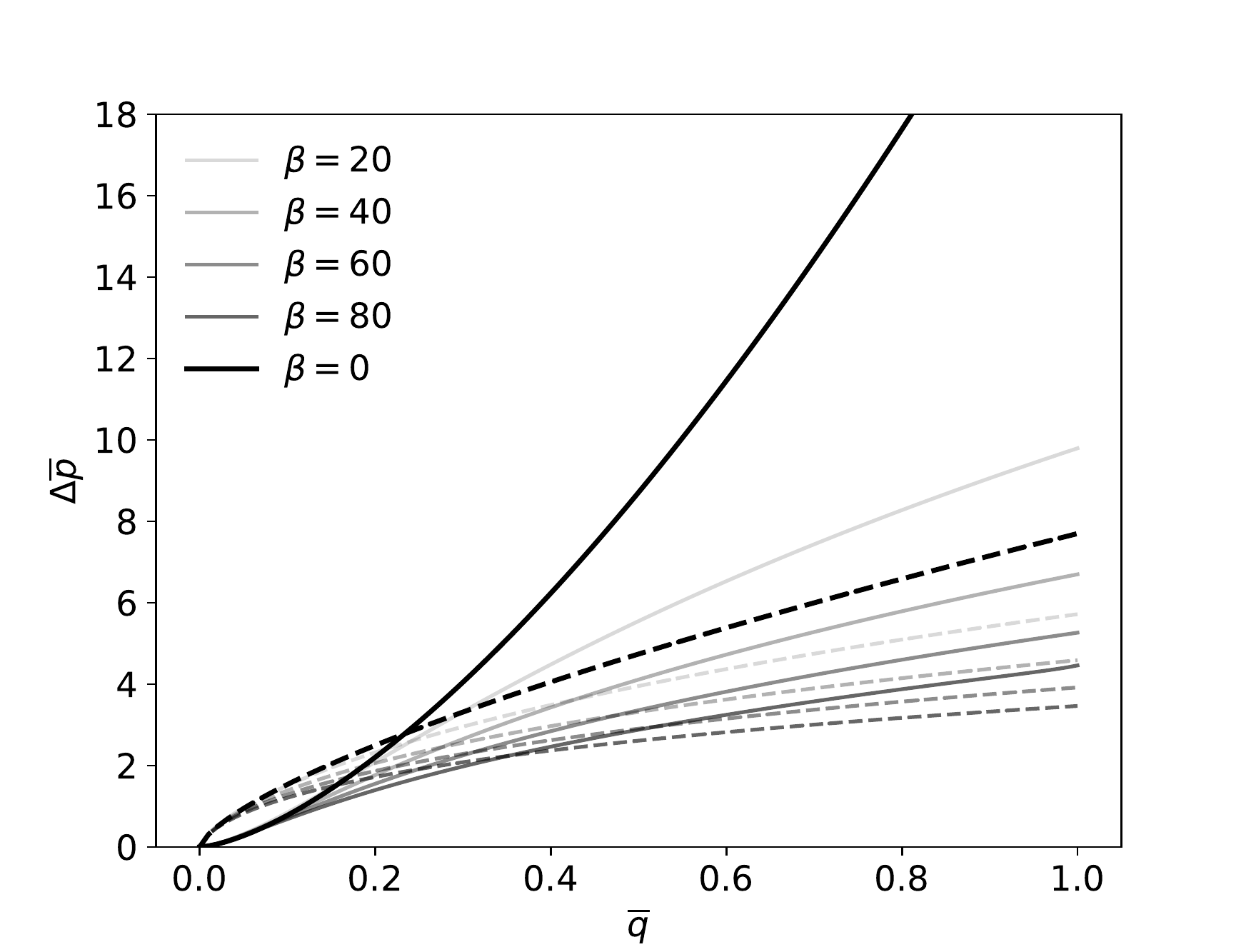}
	\caption{The dimensionless pressure drop $\Delta \bar{p}$ across the microchannel for different values of the FSI parameters $\beta$ and a fixed top wall thickness-to-width ratio of $t/w=0.2$. The dashed curves correspond to a shear-thinning fluid ($n=0.7$), while the solid curves correspond to a shear-thickening fluid ($n=1.5$).} 
\label{fig:P_Vs_Q_beta}
\end{figure}
\begin{figure}[ht!]
	\centering
 	\includegraphics[width=\linewidth]{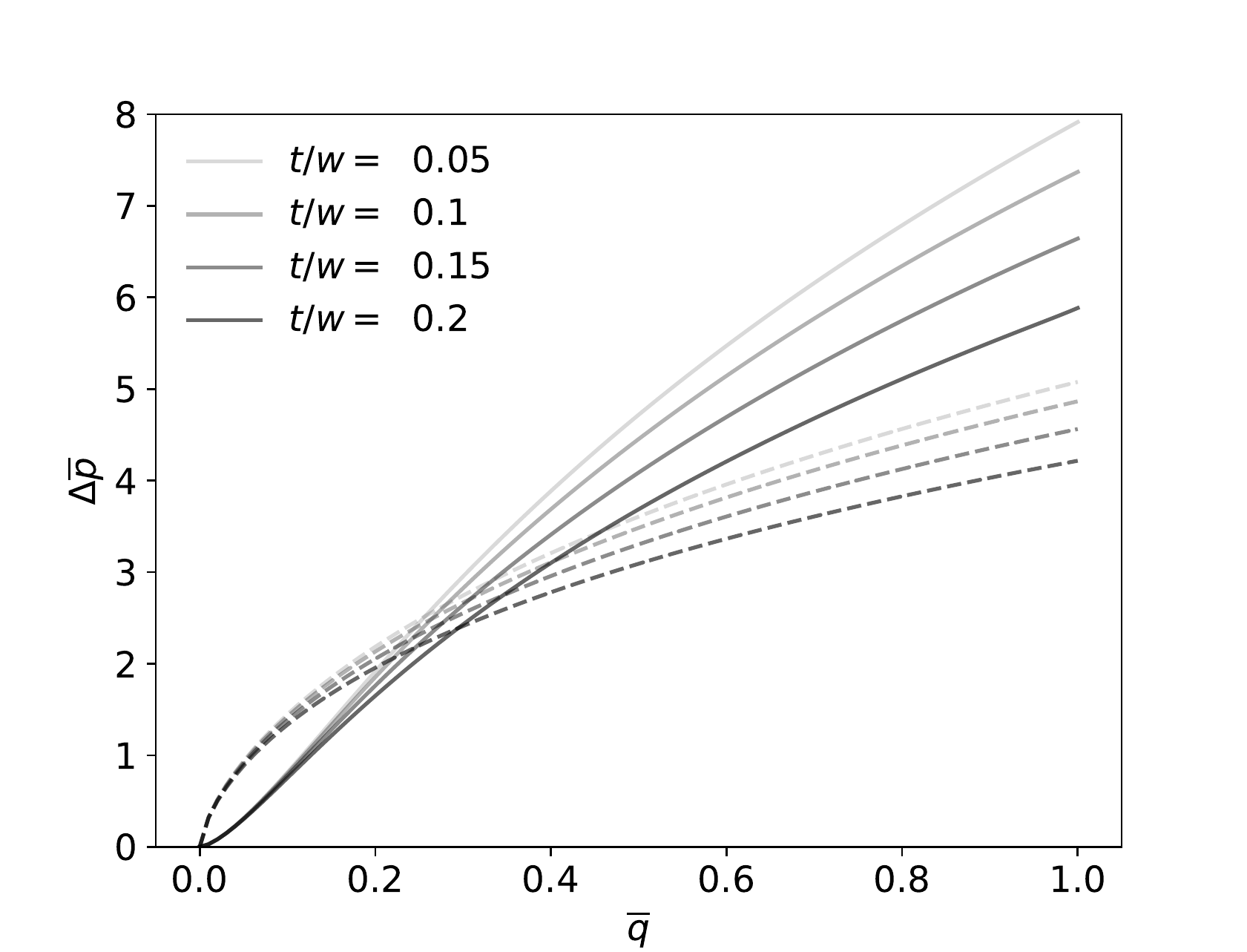}
	\caption{The dimensionless pressure drop $\Delta \bar{p}$ across the microchannel for different values of the top wall's thickness-to-width ratio $t/w$ and a fixed FSI parameter of $\beta=50$. The dashed curves correspond to a shear-thinning fluid ($n=0.7$), while the solid curves correspond to a  shear-thickening fluid ($n=1.5$).}
\label{fig:P_Vs_Q_t_W}
\end{figure}

Our mathematical analysis in Sec.~\ref{sec:finding_q_dp} culminated in Eq.~\eqref{eq:p_ode_thick_plate}, an ODE for the dimensionless pressure profile $\bar{p}(\bar{z})$. We integrate this ODE numerically using the {\tt odeint} subroutine in Python's SciPy module \cite{SciPy}. Forty terms are taken in the series in Eq.~\eqref{eq:q_thick_plate} to ensure an accurate estimate of the infinite series. It was verified that taking any more than forty terms does not influence the numerical values for the pressure thus calculated. 

In Fig.~\ref{fig:P_Vs_Q_beta}, we show the complete dimensionless pressure drop $\Delta\bar{p}$ for different values of $\beta$ but a fixed $t/w$. We note that $\Delta\bar{p}$ is larger for smaller values of $\beta$. This trend is similar to one observed in \cite{SC18,CCSS17} for Newtonian fluids. We understand that, because a large value of $\beta$ corresponds to stronger fluid--structure coupling and larger absolute deformations, the flow impedance is lower, which culminates in a smaller $\Delta\bar{p}$. 

We also see that shear-thickening fluids (solid curves) lead to a larger $\Delta\bar{p}$ than shear-thinning fluids (dashed curves) for larger $\bar{q}$, and vice versa for small $\bar{q}$. { This cross-over is due to the shear-rate-dependent apparent viscosity of power-fluids. As discussed above, the shear stress scales as $\sim m\left( {\mathcal{V}_c}/{h_0}\right)^n$ and  $\mathcal{V}_c \propto q$. Therefore, depending on whether $\mathcal{V}_c/{h_0} \gtrsim 1$ s\textsuperscript{$-1$} or $\mathcal{V}_c/{h_0} \lesssim 1$ s\textsuperscript{$-1$} ($1$ s\textsuperscript{$-1$} for the sake of this argument), the shear stress will increase or decrease with $q$, respectively. Therefore, in the first case, the shear stress in the flow of a shear-thickening fluid ($n>1$) is higher than the corresponding one in a shear-thinning fluid ($n<1$), resulting in a higher pressure drop; and, vice versa in the second case. The precise cross-over is difficult to predict due to the coupled nature of the FSI problem. For $\beta=0$, however, we can immediately predict this cross-over from Eq.~\eqref{eq:q_dpdz_beta0} by solving $[2^{1+1/n_1}(2+1/n_1)\bar{q}]^{n_1}=[2^{1+1/n_2}(2+1/n_2)\bar{q}]^{n_2}$ for $\bar{q}$ with $n_1=0.5$ and $n_2=1.5$, yielding $\bar{q}\approx 0.22964$, in perfect agreement with Fig.~\ref{fig:P_Vs_Q_beta}.}

Figure~\ref{fig:P_Vs_Q_t_W} shows $\Delta\bar{p}$ across the microchannel for different values of $t/w$ but fixed $\beta$. Clearly, $\Delta\bar{p}$ decreases with $t/w$ because a thicker top wall has a larger absolute deformation at steady state [e.g., compare Figs.~\ref{fig:Def_Axial_Thin} and \ref{fig:Def_Axial_Thick} below, and recall Eq.~\eqref{eq:Deflection_Channel2}], thus increasing the cross-sectional area, posing a lower impedance to the flow, and requiring a smaller $\Delta\bar{p}$ for the same $\beta$.

\subsection{Comparisons between theory and 3D direct numerical simulations of FSI}

In this subsection, we compare the results of our mathematical theory with those of ``full'' three-dimensional (3D) direct numerical simulations of non-Newtonian FSI in a microchannel. We carry out simulations using the commercial software suite by ANSYS \cite{ANSYS2}, building upon the success of previous computational analyses of microchannel FSIs using ANSYS' software \cite{CPFY12,SC18}. To ensure that the simulations capture all the physics of the problem, a two-way coupling approach is employed. Specifically, the two domains---fluid and solid---are separately meshed and each set of appropriate governing equations (mass and momentum conservation for the fluid and force balance for the solid) is solved separately.  A coupling module transfers information between the two solvers to ensure a fully two-way-coupled solution procedure. Although the ``full'' set of governing equations in the fluid and solid is solved, they are still assumed steady (i.e., time independent), as above. 

The governing equations of the solid mechanics problem are the force balance for a \emph{linearly} elastic continuum characterized by a Young's modulus $E$ and Poisson ratio $\nu_\mathrm{s}$; however, strains are computed from their ``full'' nonlinear (in terms of displacements) definition. The bottom and two side walls of the microchannel (see Fig.~\ref{fig:schematic}) are considered rigid (undeformable), and the solid mechanics problem for the soft (deformable) top wall is initialized with a flat rectangular plate configuration. No displacements are allowed along the planes $x=\pm w/2$ and $z=0,\ell$ to enforce clamping. For all the example cases in this subsection, the microchannel's top wall is considered to be made from PDMS, which is modeled as a nearly incompressible linearly elastic isotropic solid with $E=1.6$ MPa and $\nu_\mathrm{s}=0.499$, along the lines of \cite{SC18}.

The governing equations of the fluid mechanics problem are the (steady) Navier--Stokes equations with a variable (apparent) viscosity (i.e., the tensorial form of Eq.~\eqref{eq:constitutive}, see \cite{BAH87})  with consistency factor $m$ and power-law index $n$ to account for the assumed {power-law} rheology. To capture the flow behavior of human blood in a microchannel, we take $m = 0.018$ Pa$\cdot$s$^n$, $n=0.7$ \cite{HKP99}, which are estimates obtained by fitting blood rheology to the power-law model, and $\varrho = 1060$ kg/m$^3$ (in line with typical estimates for whole blood (see, e.g., \cite[Ch.~11]{CK98}). The fluid mechanics problem is initialized with a \emph{fully-developed} flow profile. This velocity profile is obtained by performing a separate simulation in ANSYS Fluent of just the flow in an equivalent \emph{rigid} microchannel with the same cross-section. The equivalent microchannel was chosen to be sufficiently long to allow the flow to develop fully from an initially uniform (in the cross-section) velocity with the imposed flow rate $q$. The fully developed velocity profile at the outlet of the rigid microchannel is then extracted and imposed along the inlet plane ($z=0$) of the deformable microchannel, while the pressure is set to gauge at the outlet: $p(\ell)=0$. No slip (zero velocity in this steady problem) boundary conditions are imposed on all solid and elastic channel walls. Further specific computational details and ANSYS solver settings are described in \ref{app:numerics}.

The goal of a comparison between direct numerical simulation and theory is to validate our mathematical results and to ascertain their range of validity. To explore the applicability of the theory over a wide portion of the parameter space, we separate the simulations into those suitably described by (a) {thin-plate} theory (Sec.~\ref{sec:thin}), for which the thickness-to-width ratio of the microchannel's top wall is small ($t/w = 0.075$), and (b) {thick-plate} theory (Sec.~\ref{sec:thick}), for which the thickness-to-width ratio is not as small ($t/w = 0.36$). For each of the latter cases, multiple numerical simulations are performed under the methodology just described above. The material and geometric properties are fixed but the inlet flow rate is varied from $q=7$ mL/min to $20$ mL/min for the thin-plate example and from $q=30$ mL/min to $70$ mL/min for the thick-plate example. These values were chosen to generally represent the ranges in a typical experiment (e.g., \cite{RS16,OYE13}).

\subsubsection{Thin-plate example and validation}
\label{sec:thin}

Consider a case in which the deformable top wall of the microchannel (see Fig.~\ref{fig:schematic}) is a ``thin'' plate. The geometric parameters of this model microchannel are given in Table~\ref{Table:Case1_Thin_Plate}. These values were chosen to emulate the PDMS-based microchannels (labeled ``S4'' for the thinner top wall and ``S5'' for the thicker top wall) manufactured in the experimental study \cite{OYE13}, which were also two of the examples for our Newtonian benchmark simulation study \cite{SC18}.

\begin{table}
	\centering
	\begin{tabular}{lllllll}
    \hline
	$h_0$ & $w$ & $\ell$ & $t $ & $\delta$ & $\epsilon$ & $t/w$\\ 
    \hline
    \hline
	0.244 & 2.44 & 24.4 & 0.183 & 0.1 & 0.01 & 0.075\\
    \hline
	\end{tabular}
	\caption{Dimensions and geometric parameters for the thin-plate benchmark. All lengths are given in mm.}
	\label{Table:Case1_Thin_Plate}
\end{table}

Specifically, by thin we mean that that the plate's thickness-to-width ratio is $t/w = 0.075 \ll 1$, and the plate's bending rigidity can be calculated to be $B=7.36$ $\mu$J. Furthermore, as shown in the latter table, the lubrication theory separation of scales is respected, i.e., $\epsilon \ll \delta \ll 1$, which validates the assumptions of shallowness and slenderness, respectively, imposed on the theory developed above. For the flow rates chosen, we can calculate a  maximum $\epsilon Re \approx 0.8$, which is small enough to justify the lubrication approximation. Here, $Re = \varrho\mathcal{V}_c^2/\tau_c$, $\tau_c = m (\mathcal{V}_c/h_0)^n$ is the stress scale for a power-law fluid, leading to $Re = \varrho\mathcal{V}_c^{2-n}h_0^{n}/m = \varrho q^{2-n}h_0^{2n-2}w^{n-2}/m$.

\begin{figure}
\centering\includegraphics[width=1\linewidth]{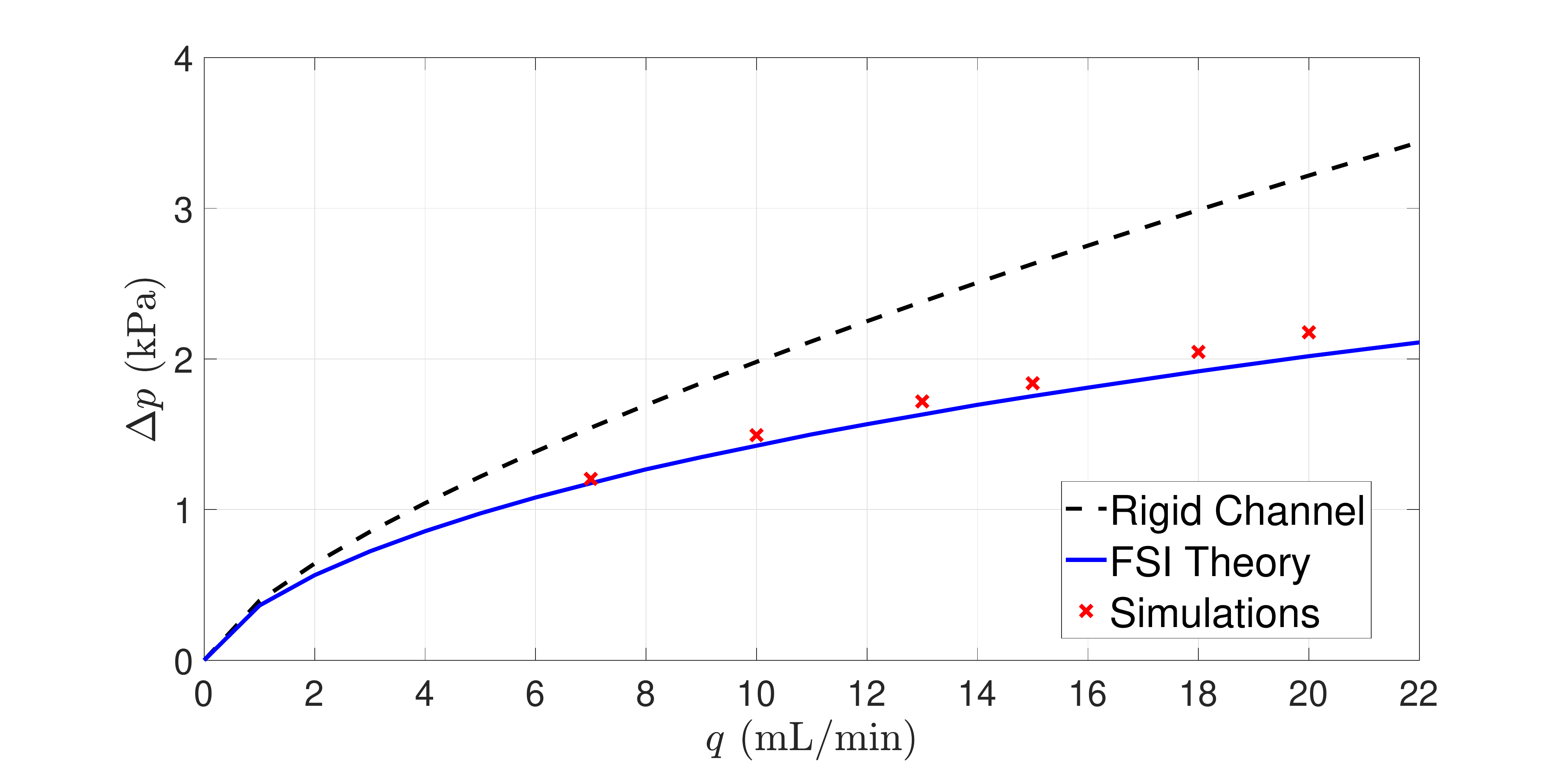}
\caption{Flow rate--pressure drop relationship for a thin-plate top wall: comparison between non-Newtonian FSI theory [i.e., $\Delta p $ computed from Eq.~\eqref{eq:p_ode_thin_plate}], direct numerical simulation, and the rigid microchannel expression [i.e., $\Delta p $ computed from Eq.~\eqref{eq:q_dpdz_beta0}].}
\label{fig:QvsP_Thin_Plate}
\end{figure}

Now, we are ready to validate our theory against direct numerical simulations. To this end, in Fig.~\ref{fig:QvsP_Thin_Plate}, we show the full dimensional pressure drop across $\Delta p$ the microchannel as a function of the imposed inlet dimensional flow rate $q$. For the range of parameters chosen, significant FSI occurs in this microchannel with a soft top wall as the differences between the ideal Hagen--Poiseuille (dashed), the theory curve (solid), and the simulation data in Fig.~\ref{fig:QvsP_Thin_Plate} clearly show. More importantly, however, there is good quantitative agreement between the theoretical prediction for the $q$--$\Delta p$ relation under FSI and the direct numerical simulation results. Overall, Fig.~\ref{fig:QvsP_Thin_Plate}, shows that our mathematical model for the $q$--$\Delta p$ relation, i.e., the solution to the ODE \eqref{eq:p_ode_thin_plate}, accurately captures the pressure drop due to non-Newtonian FSI in a microchannel with a thin top wall, within 7\% maximum pointwise error in this example. Also observe that FSI has a significant effect on the flow, reducing $\Delta p$ up to 40\% compared to a rigid channel. Clearly, it is crucial to accurately capture this quantitative difference between a deformable and a rigid microchannel when designing microfluidic systems.

\begin{figure}[t]
\centering
\includegraphics[width=1\linewidth]{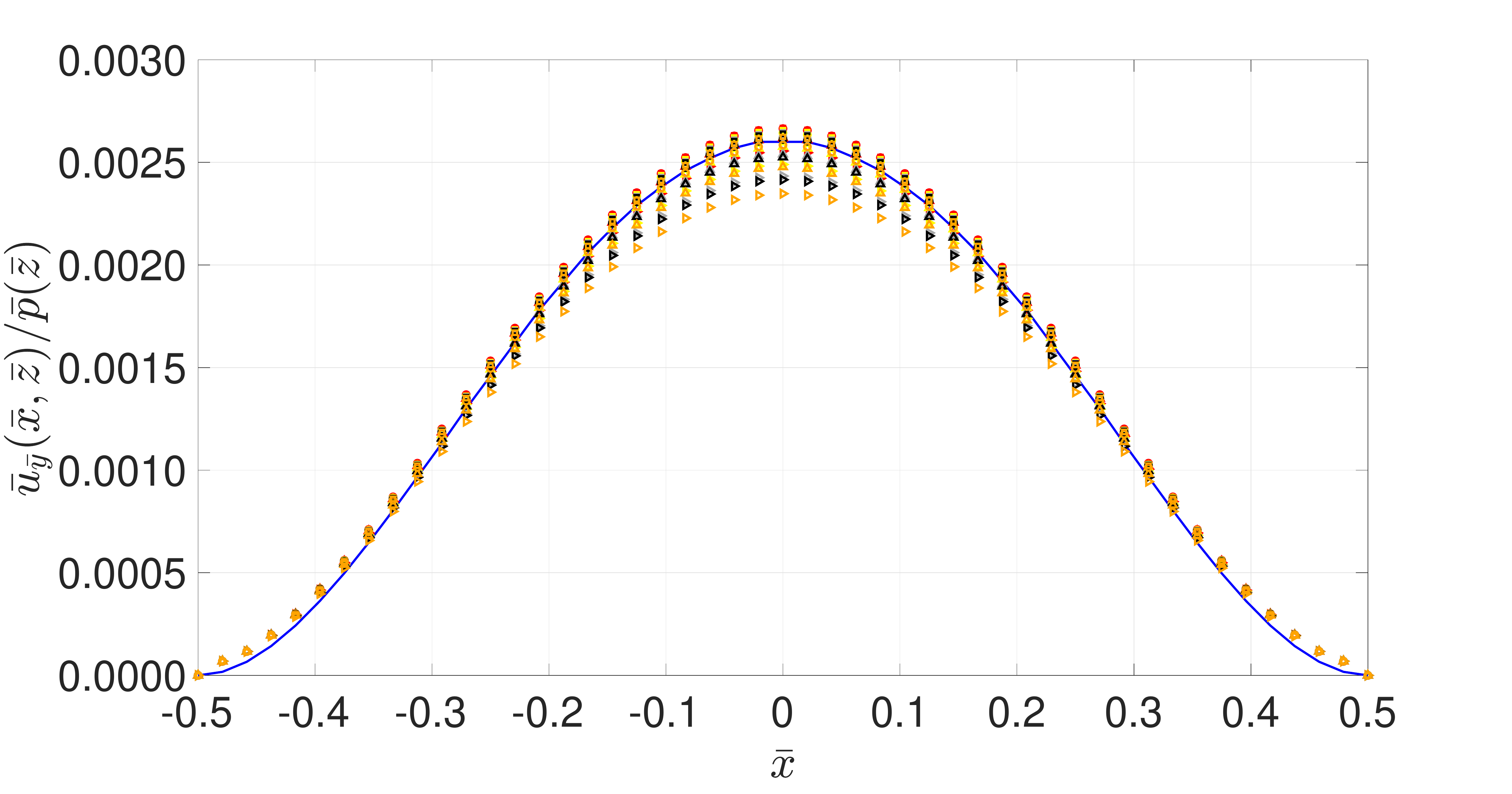}
\caption{Self-similar collapse of the scaled dimensionless cross-sectional displacement profiles, $\bar{u}_{\bar{y}}(\bar{x},\bar{z})/\bar{p}(\bar{z})$, for different flow rates of the example thin-plate case. The solid curve corresponds to the theoretical profile from Eq.~\eqref{eq:Deflection_Channel}, while the symbols correspond to direct numerical simulation. Colors: red $=7$ mL/min, yellow $=10$ mL/min, gray $=13$ mL/min, black $=15$ mL/min, and orange $=18$ mL/min; shapes: $\Box$ is $z= 21.4$ mm,  $\medcirc$ is $z=19.4$ mm, $\bigtriangleup$ is $z=16.4$ mm, and $\rhd$ is $z=12.4$ mm.}
\label{fig:Def_Axial_Thin}
\end{figure}

Switching to the structural domain, in Fig.~\ref{fig:Def_Axial_Thin}, we show the \emph{self-similar collapse} of the ratio of the dimensionless transverse deflection to the dimensionless pressure, i.e., $\bar{u}_{\bar{y}}/\bar{p}$, for different $q$ and at different axial positions $z$. This self-similar form, in which the ratio $\bar{u}_{\bar{y}}/\bar{p}$ is a \emph{universal} function of $\bar{x}$ alone is immediately apparent from the structural deformation prediction in Eq.~\eqref{eq:Deflection_Channel}. Figure~\ref{fig:Def_Axial_Thin} shows that there is good agreement between the collapsed data points and the theoretical shape predicted by Eq.~\eqref{eq:Deflection_Channel}. At higher flow rates, however, the simulation data begins to ``dip'' below the theory curve, resulting in  a maximum error of 11\%. 

\begin{figure}[t]
\centering\includegraphics[width=1\linewidth]{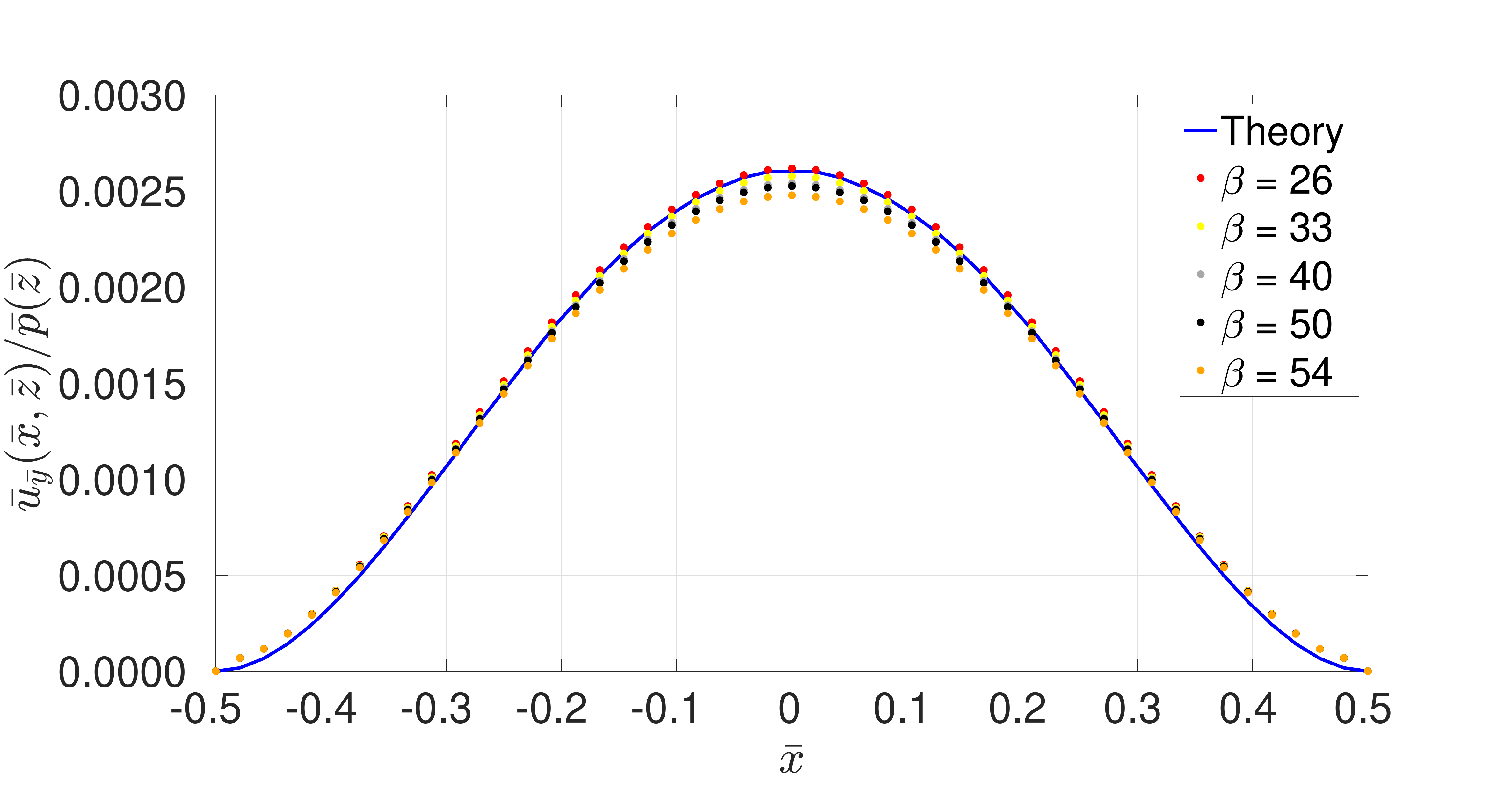}
\caption{Self-similar collapse of the scaled dimensionless cross-sectional displacement profiles, $\bar{u}_{\bar{y}}(\bar{x},\bar{z})/\bar{p}(\bar{z})$, from Fig.~\ref{fig:Def_Beta_Thin plate} for different values of the FSI parameter $\beta$ but at the fixed flow-wise cross-section  $z=16.4$ mm. Symbols correspond to the computational results from two-way FSI simulations. The solid curve corresponds to the theoretical profile from  Eq.~\eqref{eq:Deflection_Channel}.}
\label{fig:Def_Beta_Thin plate}
\end{figure}

{For a different perspective, and to further examine the collapse (or lack thereof),} in Fig.~\ref{fig:Def_Beta_Thin plate}, we plot the same data but at a \emph{fixed}-$z$ cross-section, as a function of the FSI parameter $\beta$ [recall Eq.~\eqref{eq:beta}]. Again there is good agreement between the results from direct numerical simulation and the theoretical prediction from Eq.~\eqref{eq:Deflection_Channel}, with deviations { (and the ``dip'' below the theory curve) now clearly being attributable to $\beta$ being large}.\footnote{As discussed in \cite[Appendix~B]{CCSS17}, this type of ``linear FSI'' theory holds for $\beta$ large, up to at least $\beta=\mathcal{O}(1/\delta)$.}

Therefore, we have not only shown the high accuracy of our theory through this comparison with direct numerical simulations, but we have also outlined the extent of the theory's applicability. Specifically, we observe that the systematic deviations from the theoretical prediction in Figs.~\ref{fig:Def_Axial_Thin} and \ref{fig:Def_Beta_Thin plate} occur at high $q$ and, equivalently, large $\beta$, irrespective of the axial position. This observation is attributed to the fact that at higher $q$, the deformation of the top wall enters a nonlinear stretching regime, which cannot be described by classical plate theories, as also discussed in \cite{SC18}. This hypothesis is corroborated by Fig.~\ref{fig:Def_Beta_Thin plate}, in which the collapse worsens as the FSI parameter $\beta$ increases---i.e., the load imposed onto the plate by the fluid becomes larger compared to the plate's bending rigidity. In other words, as $\beta$ increases, so does the effect of FSI, and the deformation of the top wall becomes more significant, leaving the range of validity of our ``linear'' FSI model. 

\subsubsection{Thick-plate example and validation}
\label{sec:thick}

Our benchmark microchannel for a thick-plate top wall has dimensions as given in Table~\ref{Table:Case1_Thick_Plate}. The thickness-to-width ratio for this microchannel is $t/w=0.36$, which is a large enough value to warrant the deployment of the Reissner--Mindlin thick-plate theory. This plate's bending rigidity can be calculated to be $B=266$ $\mu$J. Then, the displacement of the top wall is predicted to obey Eq.~\eqref{eq:Deflection_Channel2}. As before, to ensure the proper asymptotic regime of a long and shallow microchannel, we must have $\epsilon \ll \delta \ll 1$, which is indeed the case for this microchannel as shown in Table~\ref{Table:Case1_Thick_Plate}. 

For the flow rates chosen, we calculate a maximum $\epsilon Re \approx 8.7$, which is larger than the corresponding value for the thin-plate example above. {Although, strictly speaking, this value of $\epsilon Re$ (and not $\epsilon^2 Re$ as in two-dimensional lubrication theory \cite{CCSS17}) is beyond the expected applicability of the lubrication approximation, we nevertheless obtain good agreement between the theoretical prediction and direct numerical simulations. This highlights the extent to which lubrication theory can be ``pushed beyond'' its strict $\epsilon Re \ll 1$ validity range.} 

\begin{table}
	\centering
	\begin{tabular}{lllllll}
    \hline
	$h_0$ & $w$ & $\ell$ & $t$ & $\delta$ & $\epsilon$ &$t/w$\\ 	\hline
    \hline
	0.155 & 1.7 & 15.5 & 0.605 & 0.09 & 0.01 & 0.36\\
    \hline
	\end{tabular}
	\caption{Dimensions and geometric parameters for the thick-plate benchmark. All lengths are given in mm.}
	\label{Table:Case1_Thick_Plate}
\end{table}

\begin{figure}
\centering\includegraphics[width=1\linewidth]{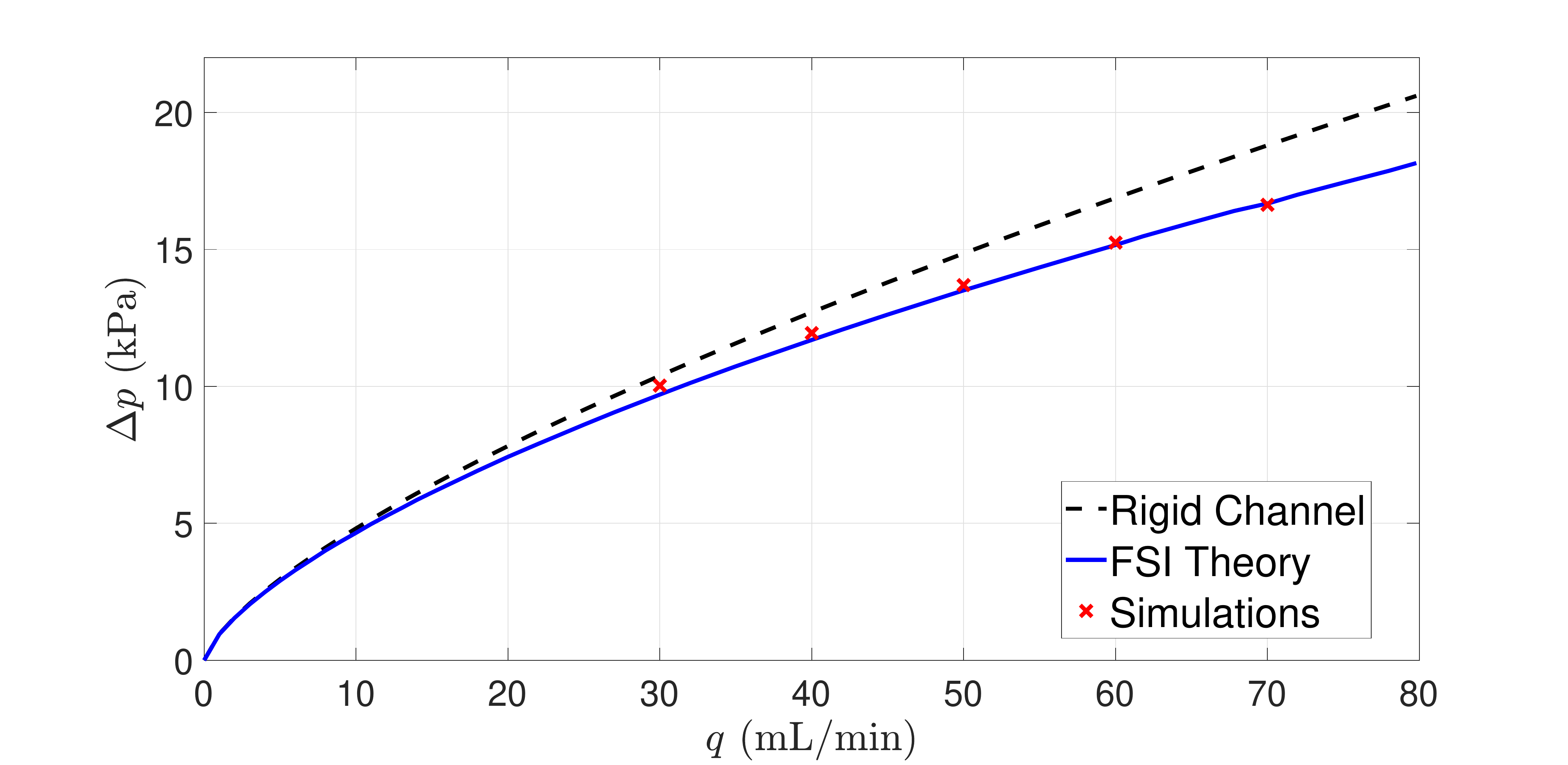}
\caption{Flow rate--pressure drop relationship for a thick-plate top wall: comparison between non-Newtonian FSI theory [i.e., $\Delta p$ computed from  Eq.~\eqref{eq:p_ode_thick_plate}], direct numerical simulations, and the rigid microchannel expression [i.e., $\Delta p $ computed from Eq.~\eqref{eq:q_dpdz_beta0}].}
\label{fig:qdp_Case1_Thick_Plate}
\end{figure}

First, we compare the flow rate--pressure drop relation predicted by our theory, i.e., the solution of Eq.~\eqref{eq:p_ode_thick_plate}, to the results of direct numerical simulation. To this end, in Fig.~\ref{fig:qdp_Case1_Thick_Plate}, we plot the pressure drop $\Delta p$ across the microchannel for different values of the imposed inlet flow rate $q$. The mathematical prediction for $\Delta p$, which is found from integrating the ODE~\eqref{eq:p_ode_thick_plate}, is in excellent agreement with the direct numerical simulation results within 2\% maximum pointwise error, across a half a decade of range in the flow rate. In this case, the pressure drops are significantly larger than in Sec.~\ref{sec:thin} because of the much smaller (undeformed) cross-sectional area (see Table~\ref{Table:Case1_Thick_Plate}). Thus, while the thick top wall deforms more than an equivalent thin one, which increases the cross-sectional area and lowers the resistance to flow, the deformation in this case is not so large as to make the pressure drops in this section comparable to those in Sec.~\ref{sec:thin}, in which the channel has larger (undeformed) cross-sectional area. Nevertheless, once again, FSI decreases $\Delta p$ in the microchannel, by up to 12\% compared to an equivalent rigid conduit, through the increase in the cross-sectional area engendered by the deformation of the top wall.

\begin{figure}
\centering
\includegraphics[width=1\linewidth]{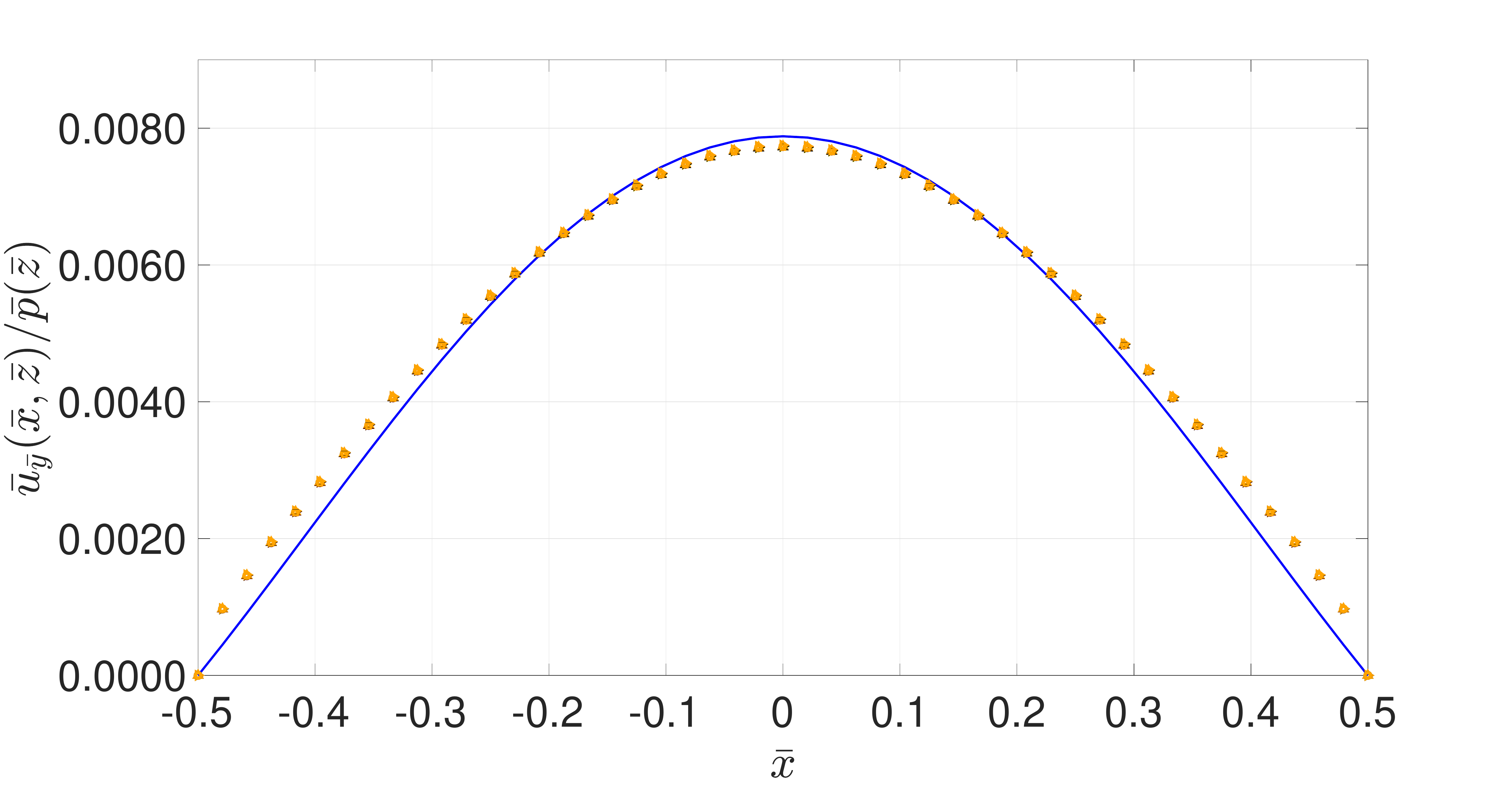}
\caption{Self-similar collapse of the scaled dimensionless cross-sectional displacement profiles, $\bar{u}_{\bar{y}}(\bar{x},\bar{z})/\bar{p}(\bar{z})$, for different flow rates of the example thick-plate case. The solid curve corresponds to the theoretical profile from Eq.~\eqref{eq:Deflection_Channel2}, while the symbols correspond to direct numerical simulation. Colors: red $=30$ mL/min, yellow  $=40$ mL/min, gray $=50$ mL/min, black $=60$ mL/min, and orange $=70$ mL/min; shapes:  $\Box$ is $z=12.5$ mm,  $\medcirc$ is $z=10.5$ mm, $\bigtriangleup$ is $z=7.5$ mm, and $\rhd$ is $z=3.5$ mm. Symbols might be hard to distinguish due to the excellent collapse of the data.}
\label{fig:Def_Axial_Thick}
\end{figure}

\begin{figure}
\centering\includegraphics[width=1\linewidth]{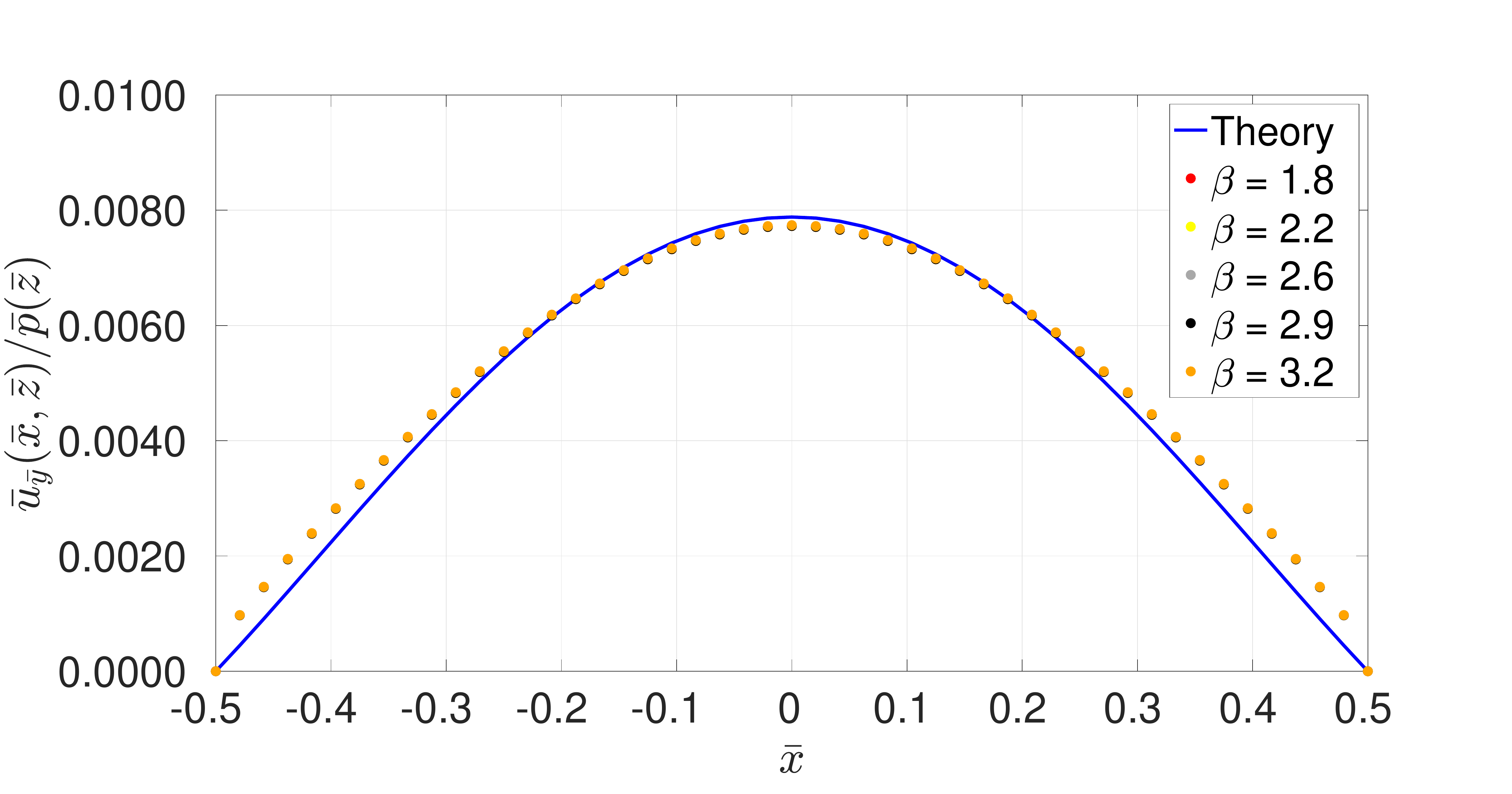}
\caption{Self-similar collapse of the scaled dimensionless cross-sectional displacement profiles, $\bar{u}_{\bar{y}}(\bar{x},\bar{z})/\bar{p}(\bar{z})$, from Fig.~\ref{fig:Def_Axial_Thick} for different values of the FSI parameter $\beta$ but at the fixed flow-wise cross-section  $z=7.5$ mm. Symbols correspond to the computational results from two-way FSI simulations. The solid curve corresponds to the theoretical profile from Eq.~\eqref{eq:Deflection_Channel2}. Symbols might be hard to distinguish due to the excellent collapse of the data.}
\label{fig:Def_Beta_Thick}
\end{figure}

Next, we plot the ratio of dimensionless deflection to dimensionless pressure, i.e., $\bar{u}_{\bar{y}}/\bar{p}$, for different values the imposed inlet flow rate $q$ in our simulations. Figure~\ref{fig:Def_Axial_Thick} shows the self-similar collapse of $\bar{u}_{\bar{y}}/\bar{p}$ across $q$ and axial positions $z$ in the flow-wise direction. There is excellent collapse of the data in Fig.~\ref{fig:Def_Axial_Thick}, with all data points being almost indistinguishable, onto the mathematically predicted dimensionless displacement profile is given by Eq.~\eqref{eq:Deflection_Channel2}. In parallel, Fig.~\ref{fig:Def_Beta_Thick} shows the same data but at a fixed value of $z$ and as a function of the FSI parameter $\beta$. Given the smaller range of $\beta$ values compared to the case in Sec.~\ref{sec:thin} (well within the ``linear'' FSI theory developed herein), we cannot visually distinguish any variation with $\beta$ in Fig.~\ref{fig:Def_Beta_Thick}.

In both Figs.~\ref{fig:Def_Axial_Thick} and \ref{fig:Def_Beta_Thick}, we observe {good} quantitative agreement between the theoretical predictions and direct numerical simulation results across the range of ${q}$, equivalently $\beta$ values chosen. {Nevertheless, there are systematic deviations between the collapsed simulation data and the $\bar{u}_{\bar{y}}/\bar{p}$ profile from Eq.~\eqref{eq:Deflection_Channel2} predicted by thick-plate theory. These deviations are due to the fact that,} in the direct numerical simulation, the 3D equations of linear elasticity are solved, while our mathematical results are based on the assumptions of the Reissner--Mindlin plate theory, which is necessarily approximate. {Specifically, the Reissner--Mindlin plate theory assumes that the normal strains, along the thickness of the plate, are small, while the shear deformations are not negligible. Therefore, every material point along the thickness of the plate undergoes the same displacement in the normal direction. This kinematic assumption may not always hold true, especially for plates that are very thick. Then, a systematic deviations from Eq.~\eqref{eq:Deflection_Channel2} (for the vertical displacements) are expected, and an even-higher-order plate theory would be required to capture them.}

\section{Conclusion}

In this paper, we presented a theory of low Reynolds number fluid--structure interactions (FSIs) between a generalized non-Newtonian fluid with shear-dependent viscosity and a rectangular microchannel with a top compliant wall. The power-law model of shear-dependent viscosity was employed for the fluid. The structural mechanics of the microchannel's top wall were modeled using two classical plate theories: the Kirchhoff--Love theory of quasi-static flexural deformations of thin plates and the first-order shear deformation Reissner--Mindlin theory applicable to thick plates. Specifically, we showed that the deformation of the structure is coupled to the fluid mechanics only through the hydrodynamic pressure, and the fluid's rheology does not explicitly affect the previously derived displacement profiles \cite{CCSS17,SC18}. Then, through a perturbative analysis under the lubrication approximation (i.e., a long and shallow microchannel), we reduced the coupled problem to a single ordinary differential equation (ODE) for the hydrodynamic pressure:
\begin{multline}
\frac{\rd{p}}{\rd{z}} = -\frac{m}{h_0}\left(\frac{q}{wh_0^2}\right)^n \left\{\frac{1}{2^{1+1/n}(2+1/n)} \sum_{k=0}^{\infty}C_{k,n} \left[\frac{w^4}{24Bh_0}p(z)\right]^k \right.\\
\times 2^{-1-4k} \left[1+\frac{8(t/w)^2}{\kappa(1-\nu_\mathrm{s})}\right]^{k} \sqrt{\pi} \, \Gamma{(k+1)}\\
\left. \times \,{}_2\tilde{F}_1\left(\frac{1}{2},-k;\frac{3}{2}+k;\frac{1}{1+\frac{8(t/w)^2}{\kappa(1-\nu_\mathrm{s})}}\right)\right\}^{-n}.
\label{eq:p_ode_thick_dimensional}
\end{multline}
Once the pressure $p(z)$ is determined via Eq.~\eqref{eq:p_ode_thick_dimensional}, the deformed channel shape is given by
\begin{equation}
h(x,z) = h_0 + \frac{w^4}{24B} \left[\frac{1}{4} - \left(\frac{x}{w}\right)^2\right] \left\{ \frac{2(t/w)^2}{\kappa(1-\nu_\mathrm{s})} + \left[\frac{1}{4} - \left(\frac{x}{w}\right)^2\right] \right\} p(z).
\label{eq:h_deform_dimensional}
\end{equation}
Equations~\eqref{eq:p_ode_thick_dimensional} and \eqref{eq:h_deform_dimensional}, technically derived under the Reissner--Mindlin plate theory, unify both the plate theories considered herein because letting $t/w\to0^+$ (recall Sec.~\ref{sec:checks2}) reduces Eqs.~\eqref{eq:p_ode_thick_dimensional} and \eqref{eq:h_deform_dimensional} to the ones derived under the Kirchhoff--Love thin-plate theory. 

Next, we illustrated some trends predicted by our theory. A microchannel with a thicker top wall is more compliant in the sense that the absolute deformation is larger for the same undeformed cross-section. Therefore, the pressure drop, at a fixed inlet flow rate, decreases with the thickness-to-width ratio $t/w$. The same trend was shown with respect to the dimensionless FSI parameter 
\begin{equation}
	\beta = \frac{w^4\mathcal{P}_c}{Bh_0},
\end{equation}
which quantifies the strength of the FSI coupling in the problem with $\beta$ large corresponding to significant compliance of the top wall. {Recall that the choice of pressure scale $\mathcal{P}_c$ was discussed after Eq.~\eqref{eq:ndvars} above.}

Full-scale direct numerical simulations of non-Newtonian {microfluidic} FSI were carried out using the commercial computer-aided engineering (CAE) platform by ANSYS \cite{ANSYS2}. The numerical simulations were used to confirm our theoretical predictions for flow and deformation, and to also ascertain the theory's range of validity. Specifically, the pressure drop across the microchannel predicted by our theory was shown to closely match the predictions of CAE simulations, across a range of flow rates, with an accuracy of 2\% in the best case and 7\% in the worst case. Deviations between theory and simulations emerge only when the structural response of the top wall of the microchannel enters a nonlinear regime ($\beta$ not small), i.e., when the deformation is too large to be adequately described within the framework of the linear strain--displacement relationships of classical plate theories. Perhaps more importantly, we note that pressure drop across the microchannel decreases significantly, up to 40\% in some cases, due to the presence of FSI. This result provides concrete new evidence for the early observations in \cite{GEGJ06} toward the fact that FSI must be accurately accounted for, when formulating flow rate--pressure drop relationships for compliant microchannels. 

Future work could include extending our results to rarefied (gas) flows at {or below that micron scale}, which brings up the issue of wall slip and compressibility \cite{EJG18}. It would also be of interest to determine the first correction due to friction at the lateral side walls, i.e., to relax the assumption $\delta \ll 1$. In \cite{CCSS17}, the latter calculation was accomplished by using the known two-dimensional (2D) solution for flow of a Newtonian fluid in a rectangular channel. For a power-law fluid, however, the rate of shear strain becomes a nonlinear function of the velocity gradient in 2D, thus an exact 2D flow solution is unavailable \cite{M65}. The latter fact makes the theoretical extension of \cite[Appendix~B]{CCSS17} a challenging open problem. Finally, it may be of interest to study fluid--structure interactions due to non-Newtonian flow in non-Cartesian geometries, continuing along the lines of the work in \cite{BBG17}. Specifically, it has been shown through 3D two-way coupled FSI simulations that the non-Newtonian nature of blood (i.e., its generalized Newtonian shear-thinning viscosity) leads to significant differences in the arterial wall deformation due to pulsatile blood flow \cite{JMS10}.


\section*{Acknowledgements}
V.A.\ and I.C.C.\ were supported, in part, by the U.S.\ National Science Foundation under grant No.\ CBET-1705637. J.D.J.R.\ was supported by the 2018 Purdue Undergraduate Research Experience (PURE) under the Purdue-India Initiative. We thank Tanmay C.\ Shidhore for comments on the manuscript.  {We thank the reviewers for their helpful comments on the manuscript.} 

\appendix

\section{Further details of the numerical simulations}
\label{app:numerics}

The governing equations of fluid mechanics and structural mechanics are solved using a finite volume and a finite element method (FEM), respectively. The finite volume solution is provided by ANSYS Fluent, while the FEM solution is provided by the ANSYS Static Structural module within ANSYS Mechanical. An iterative procedure couples the two domains/solvers and ensures that traction forces from the flow simulation are imposed onto the structural simulation as boundary conditions. Then, the displacements of the microchannel walls are calculated from the FEM analysis and transferred to the fluid problem as changes in the domain (and mesh) shape. This two-way transfer of data between the fluid and structural solvers continues till a specified (default) convergence criterion is reached. 

The displacement of the nodes on the fluid-solid interface are calculated from the FEM solution, then they are imposed onto the finite volume mesh for the fluid domain. Therefore, in each iteration of the two-way coupling loop, the fluid mesh deforms. The deformed fluid mesh requires \emph{smoothing} to preserve its overall quality, and thereby the quality of the solution to the fluid mechanics problem. Various mesh smoothing algorithms are available in ANSYS Fluent: diffusion-based smoothing, Laplace smoothing and spring-based smoothing. For FSI simulations, with hexahedral meshes, ANSYS recommends the diffusion-based smoothing algorithm. This smoothing algorithm has a parameter, denoted $\alpha$, which determines how \emph{deep} (roughly speaking) into the fluid domain mesh the effects of the deformed interface should be felt. In order to determine an optimal value of $\alpha$, we ran several test FSI simulation with different values of $\alpha$. The results are shown in Table \ref{Table:Algorithm_Testing}. For different values of $\alpha$, the pressure drop across the microchannel $\Delta p$ and the force transfers ($F_x$ in the $x$-direction, and so on) across the interface do not vary by more than $1\%$. Therefore, for our FSI simulations, the choice of $\alpha$ is not important in obtaining accurate results. Thus, we choose $\alpha = 1$ for all of our simulations reported in this paper.

\begin{table}
\centering
\begin{tabular}{lllll}
\hline
$\alpha$ & $F_x$ ($\mu$N) & $F_y$ (N) & $F_z$ (N) & $\Delta p$ (Pa) \\
\hline
\hline
$0$ & $-16.4$ & $21827$ & $295.38$ & $1622.16$\\
$1.0$ & $-15.5$ & $21813$ & $295.38$ & $1622.12$\\
$1.5$ & $-15.5$ & $21809$ & $295.39$ & $1621.99$\\ 
\hline
\end{tabular}
\caption{Force $(F_x,F_y,F_z)$ transfer and pressure drop $\Delta p$ across the microchannel for different values of the diffusion-based smoothing algorithm's parameter $\alpha$ in ANSYS Fluent.}
\label{Table:Algorithm_Testing}
\end{table}

Hexahedral elements were used for meshing the fluid and the solid domains of each benchmark case above. The aspect ratio of the elements was kept close to 1. For the case given in Table~\ref{Table:Case1_Thin_Plate}, the number of divisions of the grid was set to be 800 and 50 for the fluid domain, and 500 and 40 for the solid domain, along the length and width dimensions, respectively. Similarly, for the case given in Table~\ref{Table:Case1_Thick_Plate}, the number of divisions was set to be 1000 and 80 for the fluid domain, and 500 and 40 for the solid domain, along the length and width dimensions, respectively. The total simulation run time for the highest flow rate in each benchmark case was approximately 2 hours using 8 cores on our computational server (with, specifically, Intel Xeon E5-2680 v4 processors with 14 cores each at 2.4GHz). We also carried out various grid-independence tests, the details of which are similar to the ones reported in \cite{SC18}.

\section{The special case of a Newtonian fluid}
\label{app:coeff_match}

For a Newtonian fluid, $n=1$. Then, the summation in Eq.~\eqref{eq:channel_Q_n1_integrated} can be shown to terminate at $k= 2 + 1/n = 3$. The non-zero coefficient in the summation are:
\begin{subequations}\begin{alignat}{3}
k&=0:\quad \lim_{k\to0}  C_{k,1} \frac{k \sqrt{\pi}\, \Gamma(2 k)}{2^{4 k} \,\Gamma(3/2 + 2 k)} \;\; &=& \;\; 1, \\
k&=1:\quad \frac{C_1}{2} \frac{\sqrt{\pi}\, \Gamma(2)}{2^{4} \,\Gamma(3/2 + 2)}\left({\frac{\beta}{24}}\right)  \;\; &=& \;\; \frac{\beta}{480}, \\
k&=2:\quad \frac{C_2}{3} \frac{2 \sqrt{\pi}\, \Gamma(4)}{2^{8} \,\Gamma(3/2 + 4)}\left({\frac{\beta}{24}}\right)^2 \;\; &=& \;\; \frac{\beta^2}{362880}, \\
k&=3:\quad \frac{C_3}{4} \frac{6 \sqrt{\pi}\, \Gamma(6)}{2^{12} \,\Gamma(3/2 + 6)}\left({\frac{\beta}{24}}\right)^3 \;\; &=& \;\; \frac{\beta^3}{664215552},
\end{alignat}
\end{subequations}
as expected \cite[Eq.~(4.3)]{CCSS17}.

The non-zero coefficient in the summation in Eq.~\eqref{eq:channel_Q_thick_n1} for the thick-plate theory, in the limit of a Newtonian fluid, are calculated similarly. However, we must further show how the regularized hypergeometric function reduces to the functions denoted by $f_1$, $f_2$ and $f_3$ in \cite[Eqs.~(18)--(20)]{SC18}. To this end, consider the factor multiplying $\frac{1}{k+1}\left(\frac{\beta\bar{p}}{24}\right)^k$ in a representative term in the series in Eq.~\eqref{eq:channel_Q_thick_n1}:
\begin{multline}
C_{k,n}2^{-1-4k} \left[1+\frac{8(t/w)^2}{\kappa(1-\nu_\mathrm{s})}\right]^{k}\sqrt{\pi} \, \Gamma{(k+1)}\\ \times \,{}_2\tilde{F}_1\left(\frac{1}{2},-k;\frac{3}{2}+k;\frac{1}{1+\frac{8(t/w)^2}{\kappa(1-\nu_\mathrm{s})}}\right).
\label{eq:example_factor}
\end{multline}
For integer $k$, the expression in the last equation evaluates to a polynomial in $\xi = (t/w)^2/[\kappa(1-\nu_\mathrm{s})]$ via the properties of the regularized hypergeometric function ${}_2\tilde{F}_1$. Specifically, we may verify, e.g., using {\sc Mathematica}, that for $k=0$, Eq.~\eqref{eq:example_factor} reduces to $1$. Then for $k=1$, Eq.~\eqref{eq:example_factor} becomes
\begin{equation}
\frac{1}{10} + \xi,
\end{equation}
which is precisely the factor in the brackets multiplying $p(z)$ in \cite[Eqs.~(17)]{SC18} after $1/48$ is factored out. Likewise, for $k=1$, Eq.~\eqref{eq:example_factor} becomes
\begin{equation}
\frac{1}{210} + \frac{3}{35}\xi + \frac{2}{5} \xi^2,
\end{equation}
which is precisely the factor in the brackets multiplying $p(z)^2$ in \cite[Eqs.~(17)]{SC18} after $1/1728$ is factored out, and so on.


\bibliographystyle{elsarticle-num.bst}
\bibliography{Mendeley.bib}


\end{document}